\documentclass[onecolumn,showpacs,preprintnumbers,superscriptaddress,nofootinbib]{revtex4-1}

\usepackage{amsmath}
\usepackage{amsfonts}
\usepackage{amssymb}
\usepackage{graphicx}
\usepackage{color}
\usepackage{epstopdf}

\usepackage{booktabs}
\usepackage{hyperref}
\usepackage{cleveref}
\usepackage{color}
\usepackage{braket}
\usepackage{mathtools}
\usepackage[rightcaption]{sidecap}
\usepackage{subfigure}

\def\mnras{MNRAS}
\def\apj{ApJ}
\def\apjl{ApJL}
\def\apjs{ApJS}

\def\aap{A\&A}

\def\prd{Phys. Rev. D}
\def\apss{Astrophysics and Space Science}
\def\jcap{JCAP}
\def\nat{Nature}
\def\nar{New Astronomy Reviews}
\def\pasj{Publications of the Astronomical Society of Japan}
\def\memsai{Mem. Soc. Ast. It.}

\graphicspath{{Graphics/}}

\def\be{\begin{equation}}
\def\ee{\end{equation}}
\def\bea{\begin{eqnarray}}
\def\eea{\end{eqnarray}}

\def\mnras{MNRAS}
\def\apj{ApJ}
\def\apjl{ApJL}
\def\apjs{ApJS}

\def\aap{A\&A}

\def\aj{AJ}

\def\araa{Annual Review of A\&A}

\def\physrep{Physics Report}
\def\actaa{Acta Astronomica}

\begin{document}

\title{A roadmap to gamma-ray bursts: new developments and applications to cosmology}


\author{Orlando Luongo}
\email{orlando.luongo@unicam.it}
\affiliation{School of Science and Technology, University of Camerino, Via Madonna delle Carceri 9, 62032, Camerino, Italy.}
\affiliation{Dipartimento di Matematica, Universit\`a di Pisa, Largo B. Pontecorvo 5, Pisa, 56127, Italy.}
\affiliation{Institute of Experimental and Theoretical Physics, Al-Farabi Kazakh National University, Almaty 050040, Kazakhstan.}



\author{Marco Muccino}
\email{marco.muccino@lnf.infn.it}
\affiliation{NNLOT, Al-Farabi Kazakh National University, Al-Farabi av. 71, 050040 Almaty, Kazakhstan.}
\affiliation{Istituto Nazionale di Fisica Nucleare (INFN), Laboratori Nazionali di Frascati, 00044 Frascati, Italy.}



\begin{abstract}
Gamma-ray bursts are the most powerful explosions in the universe and are mainly placed at very large redshifts, up to $z\simeq 9$. In this short review, we first discuss gamma-ray burst classification and morphological properties. We then report the likely relations between gamma-ray bursts and other astronomical objects, such as black holes, supernovae, neutron stars, etc., discussing in detail gamma-ray burst progenitors. We classify long and short gamma-ray bursts, working out their timescales, and introduce the standard fireball model. Afterwards, we focus on direct applications of gamma-ray bursts to cosmology and underline under which conditions such sources would act as perfect standard candles if correlations between photometric and spectroscopic properties  were not jeopardized by the \emph{circularity problem}. In this respect, we underline how the shortage of low-$z$ gamma-ray bursts prevents anchor gamma-ray bursts with primary distance indicators. Moreover, we analyze in detail the most adopted gamma-ray burst correlations, highlighting their main differences. We therefore show calibration techniques, comparing such treatments with non-calibration scenarios. For completeness, we discuss the physical properties of the correlation scatters and systematics occurring during experimental computations. Finally, we develop the most recent statistical methods, star formation rate and high-redshift gamma-ray burst excess and show the most recent constraints got from experimental analyses.
\end{abstract}


\maketitle
\tableofcontents

\section{Introduction}

Gamma-ray bursts (GRBs) represent powerful extra-galactic transient that emit in $\gamma$-rays \cite{solo1,solo2a}. They are commonly associated with death of massive stars or with binary compact object mergers. As expected, due to their enormous luminosity, after the aforementioned processes, there would be the presence a newborn stellar mass black hole (BH) that provides particle accelerations and emits a relativistic collimated outflow, in the form of jets. At the same time, this new system furnishes non-thermal emissions at almost all wavelengths. The above picture lies on the standard model describing GRBs and requires isotropic energies in the range $10^{44}$--$10^{47}$~J, or $10^{51}$--$10^{54}$~erg, mostly larger than the brightest supernova (SN) emission, lying on $10^{42}$~J, or $10^{49}$~erg \cite{Koveliotou1993,vonKienlin2014}. Thereby the need of singling out GRB progenitors is essential to disclose their fundamental properties as well as the physical conditions that permit  relativistic jet to born and accelerate. Even though a clear landscape for GRB progenitor is still unclear, in view of their duration it is plausible to classify GRBs into long and short ones.

Clearly, in our \emph{Precision Cosmology} epoch GRBs could open new windows\footnote{New data come from the newly born gravitational wave, neutrino and BH astronomy. We remark that precision cosmology is essential to shed light on the mysteries that jeopardize the standard cosmological puzzle. In this respect, GRBs could play a significant role since they represent outstanding explosions whose nature can trace the dark energy (DE) and BH natures.} toward the universe description at intermediate redshifts\footnote{GRBs approximately span in the range $z\simeq2$--$10$.} \cite{solo4,solo5,Berger2005}, {\it i.e.}, much larger than SN ones \cite{Rodney2015}. Thus, several new observations have been developed, with always better accuracy, trying to standardize GRBs and to handle their emissions in analogy to SNe. In general, the most tricky challenge for cosmology is measuring  distances and arguing luminosity in the cosmic scenario, understanding from astronomical emission at which distance the emitter is placed \cite{orlando1}.

Unfortunately, this is not exactly the case of GRBs that are not standard candles, {\it i.e.}, they do not provide the above requirement on distance and luminosity \cite{2021arXiv210512692K,2015MNRAS.453..128L}. In fact, their highly variable $\gamma$-ray emission, mostly evident during the prompt phase, is thought to be associated with jet internal energy dissipation. However, the jet kinematics, among all its speed, collimation, energy, magnetization, etc., are all properties not well clarified, as well as energy dissipation mechanisms and/or shock acceleration efficiency. Hence, it is hard to relate luminosity to GRB distances as their microphysics is not well understood. Although the above caveats plague the overall GRB scenario, both short and long GRBs have relativistic outflows and share analogous properties\footnote{Like the long-lived GeV emission, which is consistent
with the afterglow emission of a blast wave in adiabatic expansion} and many attempts have been spent to standardize GRBs for both clarifying their nature, internal structure and origin together with employing these objects for cosmological purposes \cite{Demianskietal2017a,solo8}.

In this review, we first introduce the concept of GRB and their main observable quantities. As stated above, according to time
duration, we introduce the role of the $t_{90}$ duration to classify GRBs following the standard guidelines and underline the issues related to such a classification, e.g. ultra-long GRBs and X-ray flashes.
To this end, we introduce the concepts of GRB progenitor, showing quantitatively the physical reasons that limit GRBs to be fully considered as genuine standard candles. However, we also emphasize how using luminosity correlations found in prompt and afterglow phases would be useful to characterize some sort of standardization technique.
In this respect,
we portray the main observable quantities coming from GRBs and deeply introduce the standard picture of GRB formation and evolution, dubbed the \emph{fireball model}.

From all the above aspects, we expect GRBs to able to reconcile the cosmic expansion history at small and intermediate redshifts, connecting \emph{de facto} late with early times, trying to open new windows toward the comprehension of cosmology. We therefore explain how GRBs serve as  complementary probes to frame DE and cosmic expansion throughout the universe evolution, together with other standard candles, e.g. type Ia SN (SNeIa), baryonic acoustic oscillation (BAO), cosmic microwave background (CMB), Hubble differential data, etc. We show how to combine such data sets with GRBs and write the main features of experimental analysis for cosmological purposes.
Great emphasis will be devoted to the circularity problem that essentially plagues cosmology with GRBs. Once introduced, we also underline strategies that do not take into account its role for fitting cosmological models with GRBs.

Hence, we provide how to challenge the standard cosmological model, namely the $\Lambda$CDM paradigm, with GRBs. To do so, we provide the main and evident features of cosmology with GRBs by showing how to perform error analyses,  Bayesian treatments and how to handle systematics for several GRB correlations. We therefore develop model dependent and independent techniques of calibrations and report a few numerical outcomes related to GRBs, showing the most recent cosmological bounds, found with distinct procedures.

The review is split as follows. After this short introduction, in Sect. II, we classify GRBs and we report the most interesting properties, among all the classification, the progenitors and the main observable quantities coming from GRBs. In Sect. III, we work out the standard GRB model, namely the fireball paradigm. Here, we also discuss about particle and radiative processes, giving emphasis to the possible emissions coming from GRBs. In Sect. IV, we start introducing the concept of cosmology with GRBs. We thus highlight distance indicators and the concept of standard candles. In Sect. V, we explain in detail the experimental tools useful for getting Baysian analysis with GRBs. Finally, in Sect. VI, we provocatively report the concept of standardizing GRBs to permit those objects to be used in the same manner than other probes. Several issues have been raised in Sect. VII, although the likely most serious one, the circularity problem, is described in detail in Sect. VIII, where we also stress the opposite view in which one can also avoid calibration. Last but not least, we report the most recent developments of cosmology with GRBs in Sect. IX, while we conclude our journey in Sect. X with our final outlooks and perspectives of this work.

\section{GRB classification and properties} \label{sec:2}

To achieve a recognized GRB classification, the strategy is to take into account the  most relevant astronomical properties of such objects.
Thus, as the most prominent GRB component is represented by the prompt $\gamma$-ray emission, it is straightforward to use it to define GRB classes based on similarity criteria.

The prompt $\gamma$-ray emission is characterized by highly-variable and multi-peaked light curves composed of either overlapping or distinct pulses with variable duration.
The duration of these pulses spreads within a wide time range. Since the duration is not fixed \emph{a priori}, it is natural to wonder whether one can arbitrarily define a time in which the above measures can be got. Hence, it is a consolidate convention to take the total burst duration in a time interval, dubbed $t_{90}$, evaluated in the observer frame over which the $90\%$, from $5\%$ to $95\%$, of the total background-subtracted counts are experimentally detectable.

In view of such a property, one gets a plausible classification, as we report below.


\subsection{Classification: short and long GRBs}\label{sec:2.1}

The light curve analysis of the first BATSE GRB catalogue evidenced a clear bimodal distribution of the $t_{90}$ duration, separated at roughly $2$~s, and in the hardness ratio (HR), namely the ratio of the total counts of the hard $100$--$300$~keV energy band over the softer $50$--$100$~keV band \cite{1981Ap&SS..80....3M,Koveliotou1993,2012grb..book.....K}.

This leads to the widely-adopted classification into
\begin{itemize}
    \item \textit{short--hard} ($t_{90}\lesssim 2$~s) GRBs, hereafter SGRBs,
    \item \textit{long--soft} ($t_{90}\gtrsim 2$~s) GRBs, hereafter LGRBs.
\end{itemize}

\begin{figure}
\centering
\includegraphics[width=0.8\hsize]{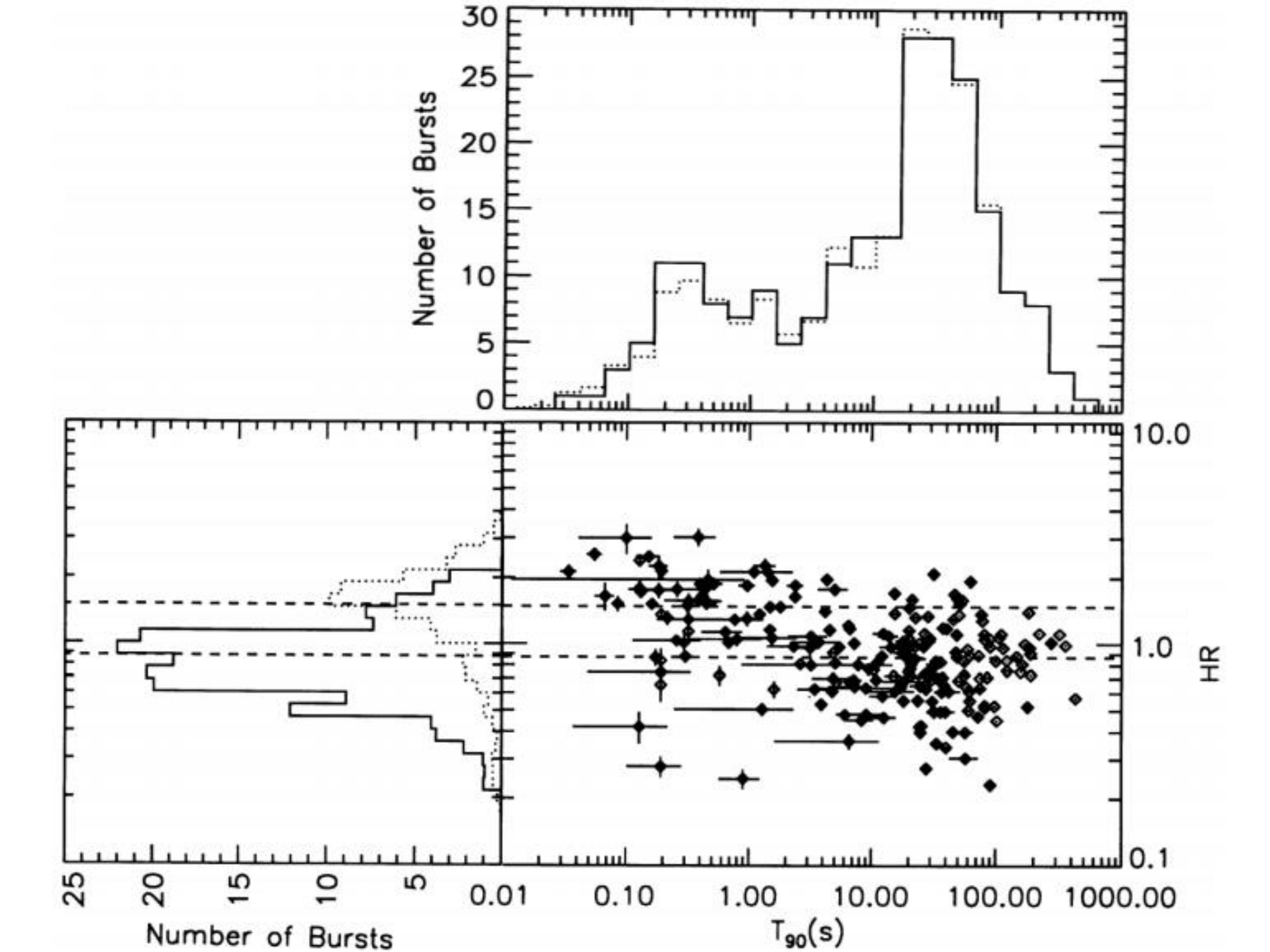}
\centering
\caption{\footnotesize GRB distribution provided by the first BATSE catalog, lying on the $t_{90}$--HR plane. The solid HR histogram shows LGRBs, whereas the dotted one is for SGRBs. The dashed horizontal lines mark mean HRs for both classes. The solid $t_{90}$ histogram represents the raw data whilst the dotted one shows the error-convolved data. Credit from Ref.~\cite{Koveliotou1993}.}
\label{fig:t90HR}
\end{figure}

The significance of such a classification scheme has been strengthened with the full $2704$ GRBs detected by BATSE and later GRB missions, providing strong evidence for two GRB progenitor channels (see e.g. Fig.~\ref{fig:t90HR}).
However, a significant overlap in the distributions of SGRBs and LGRBs suggests that {\em a more robust classification scheme based on physical properties is still missing.}

\subsection{Intermediate GRBs?}

We ended up the previous subsection, asserting the need of a more robust classification order. This scheme is veritably challenged by the existence of an intermediate class of SGRBs with extended emission (SGRBEEs), characterized by an initial short duration and spectrally-hard $\gamma$-ray pulse, followed by a softer emission lasting up to tens of seconds \cite{Norris2006,2012grb..book.....K}. Depending on the sensitivity and energy range of the GRB alert instrument and based on the above classification scheme, a SGRBEE could be classified as short or long. A GRB detector with low sensitivity at low-energy ranges in $\gamma$-ray could detect only the initial hard part of the burst (resulting in an SGRB), whereas a GRB detector with a higher sensitivity extending down to lower energies could detect also the softer extended emission (falling in the LGRB class).

A possible explanation to the origin of this extended emission involves a highly magnetized neutron star (NS) dipole spin-down emission (see Ref.~\cite{2015ApJ...805...89L} and Secs.~\ref{ssec:grbkn} and \ref{sec:magnetized_flow}).

\subsection{Ultra-long GRBs and X-ray flashes}

Further, the detection of rare events characterized by extremely long-lived prompt emissions lasting $\gtrsim10^3$~s, named ultra-long GRBs (ULGRBs), represents an additional classification threat, since it is still unclear whether ULGRBs represent a distinct class of LGRBs \cite{2014ApJ...781...13L}, or whether they are the high-end tail of the $t_{90}$ distribution of LGRBs \cite{2014ApJ...787...66Z}.

Finally, it has been reported the existence of extragalactic transient X-ray sources, dubbed X-ray flashes (XRFs), with spatial distribution, spectral and temporal characteristics similar to LGRBs \cite{Heise2001,PiranBromberg2013}.
The distinguishing properties of XRFs are

\begin{itemize}
    \item[{\bf  a)}] their observed prompt emission spectrum that peaks at energies which are an order of magnitude lower than those of standard LGRBs;
    \item[{\bf  b)}] their time integrated flux in the $2$--$30$~keV X-ray band greater than that in the $30$--$400$~keV$ \gamma$-ray band.
\end{itemize}

In view of these hazy results, classifying GRBs through $t_{90}$ and HR criteria only turns out to be puzzling, since the measured $t_{90}$ varies with energy range. Thus, the definition of a novel GRB classification scheme requires multi-wavelength criteria to better understand the physical properties behind the GRB emission.

In this respect, attempts to re-categorize GRBs from the popular long/short classes have been made in Ref.~\cite{2007ApJ...655L..25Z}, introducing alternative classes of \textit{Type I} and \textit{Type II GRBs}.
According to this scheme, the Type I class includes short/hard GRBs and SGRBEEs with no SN association, typically found in regions of their host galaxy with low star formation, and very likely originating in compact star mergers (see details in the next Sec.~\ref{progandquest}).
On the other hand, the Type II class includes long and relatively soft GRBs with SN association, usually found in star forming regions within irregular host galaxies, and thus associated with young stellar populations and likely originating in the core-collapses of massive stars (again, see details in the next Sec.~\ref{progandquest}).

Though the above scheme seems to be promising, further research on this issue is still ongoing. Therefore, for historical reasons in Sec.~\ref{progandquest} we pursue the description of the progenitor systems keeping the bimodal classification in LGRBs and SGRBs.

\subsection{Progenitors and open questions}\label{progandquest}

Beside the above discussion, the working definition of LGRBs and SGRBs suggests the existence of two different progenitor channels. Summarizing,

\begin{itemize}
    \item[{\bf  I:}] LGRBs could arise from the core-collapse of a massive star or \textit{collapsar} \cite{2006ARA&A..44..507W}, \item[{\bf  II:}] SGRBs could originate from the binary neutron star-black hole (NS-BH) or NS-NS mergers \cite{Bergerreview}.
\end{itemize}

The huge observed isotropic equivalent energy release of $\sim10^{49}$--$10^{55}$~erg implies that: for LGRBs, up to $\sim10$~M$_\odot$ are converted into radiation during the prompt emission duration of $\sim100$~s, whereas for SGRBs up to $\sim 1$~M$_\odot$ are converted into radiation within $\sim 1$~s \cite{Piran2009}.
The energy reservoir and the efficiency of the involved physical processes in producing the emitted energy represent a stringent requirement, specially for LGRBs.\footnote{For a different perspective, see e.g.~\cite{PhysRep}}.

The commonly called \textit{jets} substantially alleviate this issue by reducing the GRB energy release by  jet's correction factor $f=1-\cos\theta$. Jets can be thought, in an oversimplified picture, as outflows of relativistic matter ejected into a double-cone structure of opening angle $\theta$.
In general the jet correction is poorly constrained because requires very challenging measurements of $\theta$ and the observer viewing angle relative to the jet axis.
This makes troublesome to distinguish between geometric and dynamical effects.
Indeed, very soft GRBs could be bursts viewed off-axis, whereas low luminosity GRBs may be the result of large jet opening angles \cite{2012grb..book.....K}.

Measurements of $\theta$ can be obtained by the predicted signature of the achromatic \textit{jet breaks}, observable in the afterglow light curve at all frequencies.
This feature can be explained by the dynamics of the GRB ejecta as follows.
At the beginning, at high  values of the bulk Lorentz $\Gamma$ factor\footnote{Hereafter the bulk Lorentz factor is indicated with $\Gamma$ to avoid confusion with the power-law photon index $\gamma$, describing simple power-law GRB spectra.}, the ejecta is narrowly beamed into the jets while its Lorentz factor is $\Gamma^{-1}<\theta$ and, regardless the hydrodynamic evolution, a GRB is observed only from a small fraction of the ejecta \cite{2012grb..book.....K}.
As the ejecta decelerates, $\Gamma$ decreases below $\theta^{-1}$, the beaming angle becomes larger, and a larger portion of the ejecta becomes observable.
Continuous deceleration leads to the point that the entire surface of the jet is observable and the jet begins to spread sideways, producing a break in the light curve across the entire afterglow spectrum \cite{PanaitescuMeszaros,Sarietal1999}.
The sharpness of this break and the change in the afterglow decay rate depend on how long the jet remains collimated and on the jet radial density profile and energy distribution \cite{2009ApJ...698...43R,2021MNRAS.506.4163L}.
The time of the jet break is related to the jet opening angle, the bulk Lorentz factor and the density of the circumburst medium  (CBM).
The above description has two effects:
\begin{itemize}
    \item an  ``\textit{on-axis}'' observer detects the prompt emission and then, as the jet decelerates, the afterglow emission and finally a \textit{jet-break} due to the faster spreading of the emitted radiation;
\item an  ``\textit{off-axis}'' observer cannot detect the prompt emission, but detects an \textit{orphan afterglow}, namely an afterglow without a preceding GRB.
\end{itemize}

In the pre-\textit{Swift} era, simultaneous breaks in the optical and near-infrared (NIR) afterglow light curves were frequently interpreted as jet breaks.
Nevertheless, the improved temporal and spectral coverage of GRB afterglows, specially in X-rays by \textit{Swift}, have revealed within the first few hours after the prompt emission a complex structure made of flares, plateaus and chromatic breaks \cite{Nousek2006,Panaitescu2006,Zhang2006,Chincarini2007}.
The detected achromatic breaks are observed in a few cases.
The absence of jet-break signatures in most GRB afterglows has been interpreted as due to the over-simplified assumption homogenous jets with sharp edges, whereas more complex models now include structured jets that produce several chromatic jet-breaks, or much smoother breaks, or jets that can keep their structure for longer than previously thought making difficult to detect breaks without a wide temporal coverage \cite{2021MNRAS.506.4163L}.

Besides the jet modeling issue, any GRB model has to deal with features like very luminous X-ray flares occurring up to a few $10^4$~s after the GRB trigger and with shape and spectra similar to those flares observed during prompt emission and extended plateau phases that last for a few hours during the early afterglow evolution \cite{Zhang2006,Chincarini2007}.
Both features imply an extended central-engine activity with a continuous source of energy injection lasting the above  $10^4$~s.
In the standard picture, such long-lived energy injection requires the accretion of a significant mass onto the central BH via very large ($\sim1$~M$_\odot$) and low-viscosity ($\alpha < 10^{-2}$) accretion disk formed at the core collapse time, or via fall-back material continuously replenishing the accretion disk \cite{2008MNRAS.388.1729K}.


\subsubsection{The LGRB-supernova connection}\label{ssec:grbsn}

The possible connection between LGRB and massive progenitor stars has been speculated long before the first afterglow detection \cite{2006ARA&A..44..507W,2013RSPTA.37120275H}.
The first observational evidence came with the association between the broad line Type Ic (Ic-BL) SN~1998bw and the low-luminous LGRB~980425 at $z=0.0085$ and with lack of an optical afterglow \cite{Galama1998}.
Later on, this association was also confirmed between the Type Ic-BL SN~2003dh, temporally and spatially coincident with the standard more luminous long GRB~030329 at $z=0.1685$, with an optical afterglow light curve comparable with other cosmological GRBs \cite{Hjorth2003}.

The launch of \textit{Swift} has increased the sample of GRB-SN pairs, both spectroscopic, at $z\lesssim 0.5$ and most of them with isotropic-equivalent $\gamma$-ray energies $E_{\rm iso}<10^{49}$~erg, and photometric, in the form of SN bumps appearing in the optical afterglows $10-30$ days\footnote{In the observer frame.} after the GRB, at $z\gtrsim 0.5$ and $E_{\rm iso}\approx 10^{51}-10^{52}$~erg \cite{2013RSPTA.37120275H}. Most of the GRB-SN pairs belong to this second kind, very likely at the hand of a selection effect: the more common  low-luminosity LGRBs per unit volume are not detectable at high redshift, whereas luminous LGRBs, with higher detectability at high redshift are observed from a larger volumetric area \cite{Soderberg2006Nature}.

SNe Ic associated to some long GRBs are characterized by no hydrogen (H) and no  weak helium (He) lines \cite{2013RSPTA.37120275H}. Their occurrence close to star-forming regions offer very strong evidences that long GRBs could be associated with massive star death \cite{2013RSPTA.37120275H}.
In this regard, the best progenitor candidates are the Wolf-Rayet stars, very massive stars with an hydrogen envelope largely depleted, endowed with a fast rotation \citep{MacFadyen2001,2006ARA&A..44..507W}.
Within the collapsar model, very massive stars are able to fuse material in their centers all the way to iron (Fe).
At this point they cannot continue to generate energy through fusion and collapse mechanisms forming a BH.
Matter from the star around the core rains down towards the center and swirls into a high-density accretion disk.
In this picture, the core carries high angular momentum to form a pair of relativistic jets
out along the rotational axis where the matter density is much lower than in the accretion disk.
Jets propagate through the stellar envelope at velocities approaching the speed of light, creating a relativistic shock wave at the front \citep{Zhang2003,2012grb..book.....K}.
If the star is not surrounded by a thick,  diffuse H envelope, the leading shock accelerates as the stellar matter density decreases. Thus, by the time it reaches the star surface, $\Gamma\geq100$ is attained and the energy is released in the form of $\gamma$-ray photons \citep{Zhang2003,2012grb..book.....K}.

The collapsar model attempts to explain the time structure of GRBs prompt emission, through the modulation of the jets by their interaction with the surrounding medium, which could produce the variable Lorentz factor needful for internal shock occurrence \citep{2006ARA&A..44..507W}.
As the relativistic jet propagation through the stellar envelope of a collapsing star proceeds, its collimation was shown to occur analytically and numerically \cite{2012grb..book.....K}.
Another prediction of this model is the prolonged activity of the central engine which can potentially contribute to the GRB afterglow \citep{MacFadyen2001,Zhang2003,2006ARA&A..44..507W}.
This occurs because the jet and the disk are inefficient at ejecting all the matter in the equatorial plane of the pre-collapse star and some continues to fall back and accrete \citep{MacFadyen2001,Zhang2003,2006ARA&A..44..507W}.

The SNe associated with LGRBs appear to belong to the bright tail of  type Ic SNe and can be considered as a ``subclass'' of SNe Ic, alternatively addressed as {\em hypernovae}, in order to emphasize the extremely high energy involved in these explosions.
Remarkably, the SNe associated with both low- and high-luminous (XRFs and normal LGRBs, respectively) share very similar spectra and their peak luminosities span only $2$ orders of magnitude, whereas the associated GRBs isotropic luminosities span six orders of magnitude \cite{2013RSPTA.37120275H}.
Another distinctive feature of the GRB-SN pairs is the high photospheric expansion velocity, up to $0.1c$ \cite{2013RSPTA.37120275H}. In this scenario, one has to fit also the class of ULGRBs. The spectroscopic detection of the SN~2011kl coincident with the ULGRB~111209A \cite{2015Natur.523..189G} favors a common core-collapse origin for LGRBs and ULGRBs.
This SN exhibited a peculiar, very blue and featureless spectral shape, which was unlike other SNe Ic associated with LGRBs, but more alike the newly discovered class of superluminous SNe \cite{Gal-Yam2012}.
Other ULGRBs have either been too far or too dust-extincted to secure any detection of an underlying SN, whereas other cases evidenced the indicative flattening from a rising SN in their optical and NIR light curves at $10-20$ days after the GRB trigger \cite{2013RSPTA.37120275H}.

In this picture, however, exceptions to the LGRB-SN association have been found from deep optical observations in two nearby bursts, GRB~060505 and GRB~060614, for which the hypothetical accompanying SN would have been a hundred times fainter than SN~1998bw \cite{Fynbo2006,DellaValle2006,Gal-Yam2006}.

To conclude, ULGRBs and SNless LGRBs give evidences for the existence of further progenitor channels for LGRBs.

\subsubsection{SGRBs, macronovae and gravitational waves}\label{ssec:grbkn}

The \textit{Swift} satellite has enabled rapid and precise localisations and an increase in the number of X-ray and optical afterglow detection of both LGRBs and SGRBs.
However, SGRBs have less luminous afterglows than those of LGRBs and this fact makes difficult to obtain optical spectra and a precise burst location to plan optical follow-up to search for host galaxy associations.
The lack of any associated core-collapse SNe, the typically large offsets of the GRB position with respect to galaxy center, and the frequent association with galaxies with no ongoing star formation, provide evidences in support of a compact binary merger progenitor scenario \cite{Gehrels2005}.

The proposed progenitors for SGRBs are NS--NS and/or NS--BH binary mergers \cite{Goodman1986,Paczynski1986,Eichler1989,Bergerreview}.
These mergers take place as binary orbits decay due to gravitational radiation emission \citep{Taylor1989}.
A merger releases $5\times10^{53}$ erg, but most of this energy is due to low energy neutrinos and gravitational waves.
So, there is enough  energy available to produce a GRB, notwithstanding how a merger generates the relativistic wind required to power a burst is still object of speculations and not-well understood.
It has been argued that about one over thousand of these neutrinos annihilates and produces pairs that in turn produces $\gamma$-rays via $\nu\bar{\nu}\rightarrow e^+e^-\rightarrow\gamma\gamma$, but it has been pointed out that a large fraction of the neutrinos would be swallowed by the newly-born BH \cite{2012grb..book.....K}.

A further confirmation to the binary merger scenario consists in the detection of the so-called \textit{macronova} (MN).
The MN emission originates in NS-BH or NS-NS mergers from a fast-moving, rapidly-cooling ejected debris of neutron-rich radioactive species that decay to form transient emission and create atomic nuclei heavier than iron through neutron capture process, named {\em r-process} \cite{Li1998}.
The opacities of these produced heavy elements lead to a dim MN emission, requiring deep follow-up observations down to NIR bands.
The first indication of a MN, in the form of a re-brightening detected approximately 9~days after the GRB trigger, has been obtained by extensive follow-up of the SGRB~130603B, one of the nearest and brightest SGRBs ever detected \cite{2013Natur.500..547T}. An MN emission accompanies also the nearest SGRB ever detected, SGRB~160821B \cite{2019ApJ...883...48L,2019MNRAS.489.2104T}, and the recently detected SGRB~200522A \cite{2021ApJ...906..127F}.
For a list of other MN emissions, see Ref.~\cite{2018ApJ...860...62G}.

In the binary merger scenario, SGRBs are expected to be significant sources of gravitational waves (GWs). The smoking gun occurred on $17$ August $2017$, when the Advanced LIGO and Virgo detectors observed the event GW~170817, unambiguously detected in spatial and temporal coincident with the SGRB~170817A independently measured by the {\em Fermi} Gamma-ray Burst Monitor, and the Anti-Coincidence Shield for the Spectrometer for the International Gamma-Ray Astrophysics Laboratory \cite{ShortGW}\footnote{The GW signal, originating from the shell elliptical galaxy NGC 4993, had a duration of $\sim100$~s. By the characteristics in intensity and frequency, GW~170817 has been unambiguously associated with the inspiraling of a binary NS-SN merger of total mass $2.82_{-0.47}^{+0.09}$~M$_\odot$, which is consistent with the masses of all known binary NS systems.}.

As a further confirmation on the nature of the progenitor system of SGRB~170817A, an intense observing campaign from radio to X-ray wavelengths over the following days and weeks after the trigger led to the spectroscopic identification of a MN emission, dubbed AT~2017gfo \cite{2018NatCo...9.4089T}.

The observation of SGRB, GW and MN emission has improved our understanding of the physical properties related to the binary merger, such as the mass of the compact object, the ejected mass, and the details of the CBM surrounding the merger site.

\subsection{Observable quantities from GRBs}\label{sec:2.5}

Understanding GRB physics passes through the experimental evidence of the energy that can be collected from detectors. In particular, we can start  discussing about GRB prompt emission. It is typically observed in the hard-X (above $\sim5$~keV) and $\gamma$-ray energy domain.

\begin{figure}[t]
\centering
\includegraphics[width=0.8\hsize,clip]{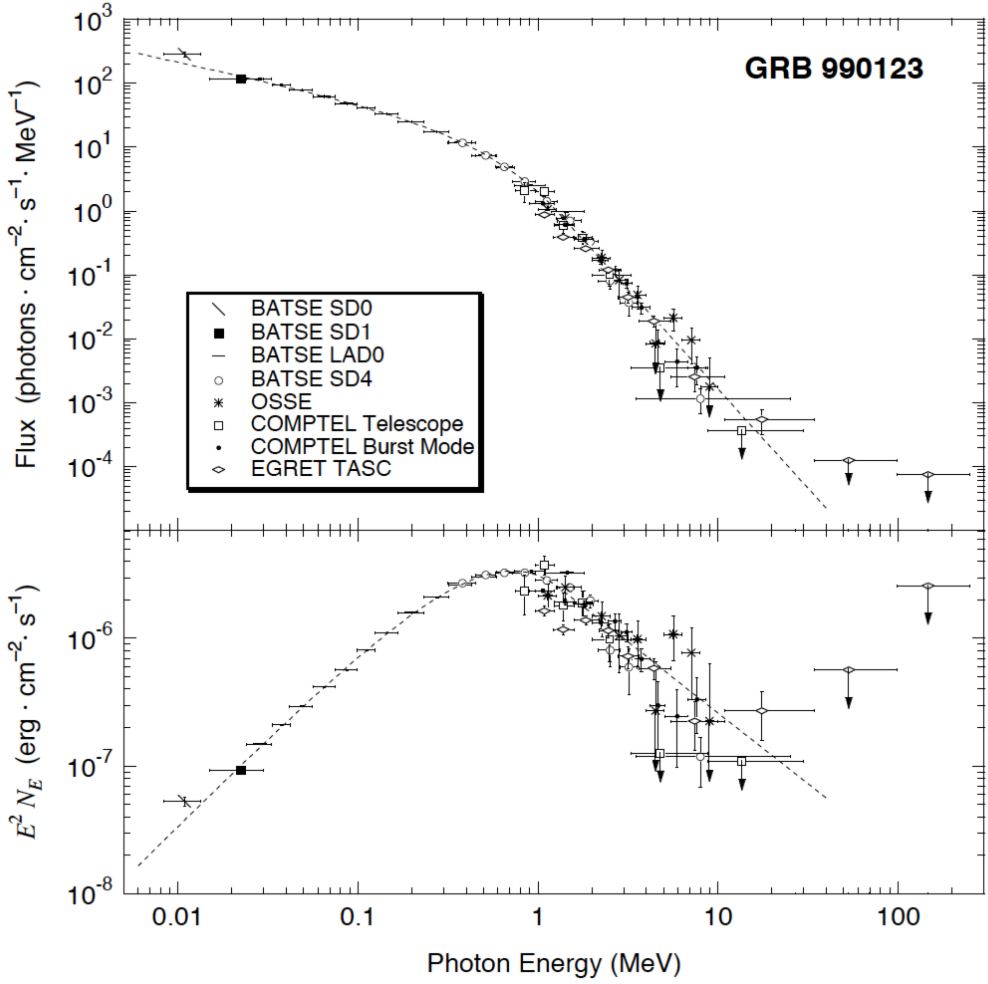}
\caption{\footnotesize Band spectral model applied to the data of GRB 990123. In the upper panel is shown the photon spectrum; in the lower panel the $E^2 N_{\rm E}$ (or $E F_{\rm E}$) spectrum. Courtesy from Ref.~\cite{Band1993}.}
\label{figBand}
\end{figure}

The operative duration of the prompt emission is due to the previously defined $t_{90}$.
Within this time interval, and also within any sub-interval with enough photons to perform a significant analysis\footnote{Typically dubbed time-integrated and time-resolved analyses, respectively.}, the observed spectral energy distribution (SED) of GRBs is non-thermal, and it is best fitted by a phenomenological model composed of a smoothly joined broken power-law called {\em Band} model \citep{Band1993} (see Fig.~\ref{figBand}).
Its functional form is
\begin{equation}
\label{Bandf}
N_{\rm E}(E)= K \left\{
\begin{array}{ll}
\displaystyle \left(\frac{E}{100}\right)^\alpha\exp\left[\frac{(2+\alpha)E}{E_{\rm p}^{\rm obs}}\right] & ,\ \ \displaystyle E\leq \left(\frac{\alpha-\beta}{2+\alpha}\right)E_{\rm p}^{\rm obs}\\
\,\\
\displaystyle \left(\frac{E}{100}\right)^\beta\exp\left(\beta-\alpha\right)\left[\frac{(\alpha-\beta)E_{\rm p}^{\rm obs}}{(2+\alpha)}\right]^{\alpha-\beta} & ,\ \ \displaystyle E>\left(\frac{\alpha-\beta}{2+\alpha}\right)E_{\rm p}^{\rm obs}
\end{array}\right.
\end{equation}
where typical  power-law index values are $-1.5\lesssim\alpha\lesssim0$ (with an average  $\langle \alpha \rangle \simeq -1$) and $-2.5\leq\beta\leq-1.5$ (with an average  $\langle \beta \rangle \simeq -2$), while the peak energy at the maximum of the of the $E^2N_{\rm E}$ (or $EF_{\rm E}$) spectrum lies within $100$~keV$\leq E_{\rm p}^{\rm obs}\lesssim$few~MeV (with an average of $\langle E_{\rm p}^{\rm obs}\rangle \simeq 200$~keV). Finally, $K$ is the normalization constant with units of ${\rm photons}\,{\rm cm}^{-2}{\rm s}^{-1}{\rm keV}^{-1}$.
In some cases the SED is also best fitted by a
power-law model\footnote{However, in this case the spectral break is very likely below or above the detector bandpass.} or by a power-law plus an exponential cutoff.
However, these models are purely mathematical, i.e. not yet physically linked to GRB intrinsic properties.
Hence, fitting data with them does not provide any insight about the emission physical origin, but may be useful for the classification scheme of GRBs and for comparing the fitted results with the predictions of different theoretical
models.

In the recent years, with a much broader spectral coverage enabled by detectors such as Fermi, evidences for more complicated broad band spectra fitted by a combined Band+thermal model, have been found in an increasing number of bursts \cite{Ryde2010,Ryde2011,Guiriec2011,2011MNRAS.418L.109G}, where the peak of the thermal component is always observed below $E_{\rm p}^{\rm obs}$.

However, the search of the best-fit model in describing GRB prompt emission spectra depends on the analysis method.
Typically, a significant spectral analysis is performed when enough photons are collected. For weak bursts, only time-integrated spectral analyses can be done and this implies that important time-dependent features may be lost or averaged, leading to a wrong theoretical interpretation.
Another issue is that the chosen spectral model is convolved with the detector response and, because of the non-linearity of the detector response matrix, this procedure cannot be inverted.
Therefore, two different models can equally provide a similar minimal difference between the model and the detected counts spectrum and lead to different theoretical interpretations.

From the fit of the time-integrated prompt emission spectrum, one can get the flux $F$ (in units of ${\rm erg}\,{\rm cm}^{-2}{\rm s}^{-1}$) on a detector energy bandpass $E_{\rm min}$--$E_{\rm max}$ as
\begin{equation}
F = \kappa \int_{E_{\rm min}}^{E_{\rm max}}{E N_{\rm E}(E) dE}
\ , \label{eq: defsbolo1}
\end{equation}
where $\kappa$ is a constant, commonly used to convert the energy, expressed in keV, to erg.

To compute the total energy emitted by a GRB in all wavelengths, a bolometric spectrum is needed.
However, the GRB prompt emission triggers $\gamma$-rays detectors in a given energy bandpass, therefore a limited part of the spectrum is available, instead of a bolometric one.
Moreover, GRBs are cosmological sources spread over a wide redshift range, so for GRBs observed by the same detector, the measured energy range corresponds to different energy bands in the cosmological rest frame of the sources.

To standardize all GRBs, fluxes are computed in the fixed rest-frame band $1$--$10^4$~keV, which is a range larger than that of most of the $\gamma$-ray detectors.
The ``bolometric'' time-integrated flux is then given by
\begin{equation}
F_{\rm bolo} = F \times \frac{\int_{1/(1 + z)}^{10^4/(1 +
z)}{E N_{\rm E}(E) dE}} {\int_{E_{\rm min}}^{E_{\rm max}}{E N_{\rm E}(E) dE}}
\ , \label{eq: defsbolo}
\end{equation}
and the total isotropically-emitted energy and luminosity are, respectively
\begin{eqnarray}
E_{\rm iso}&=&4\pi d^2_{L}F_{\rm bolo}t_{90}(1+z)^{-1}\\
L_{\rm iso}&=&4\pi d^2_{L}F_{\rm bolo}
\label{egdl}
\end{eqnarray}
where the factor $(1+z)^{-1}$ corrects the $t_{90}$ duration from the observer frame to the GRB cosmological rest-frame.
In a similar way, the peak luminosity $L_{\rm p}$, computed from the observed peak flux $F_{\rm p}$ within the time interval of $1$~s around the most intense peak of the burst light curve and in the rest frame $30$--$10^4$~keV energy band\footnote{This energy band is the one established in the original work by Ref.~\cite{Yonetoku2004}.} is given by
\begin{equation}
L_{\rm p}=4\pi d_{\rm L}^2 F_{\rm p}\,.
\end{equation}

The luminosity distance $d_{\rm L}$ depends upon the cosmological models adopted as backgrounds and can be related to the continuity equation recast as
\begin{equation}\label{lej2}
\frac{d\rho}{dz}=3\left(\frac{P +\rho}{1+z}\right),
\end{equation}
that relates the total energy density $\rho$ and pressure $P$ to the barotropic factor $\omega(z)\equiv P/\rho$ of a given cosmological model.
For a two component flat background cosmology composed of standard pressure-less matter with $\omega=0$ and a generic DE component with $\omega(z)$ (dubbed generically $\omega$CDM), the luminosity distance is then given by\footnote{Additional details on $d_L(z)$ will be summarized later in this review. Here we stress that this definition has been written for spatially flat DE models.}
\begin{eqnarray}
 d_{\rm L}(z) = (1+z) \frac{c}{H_0} \int_0^z \frac{dz'}{\sqrt{\Omega_{\rm m}(1+z')^3 + \Omega_{\rm x}f_{\rm x}(z')}} \; ,
 \label{dlum1}
\end{eqnarray}
where $H_0$ is the Hubble constant, $\Omega_{\rm m}$ and $\Omega_{\rm x}$ are the cosmological density parameters of matter and DE, respectively, and $f_{\rm x}(z)$ is given by
\begin{eqnarray}
  \label{eq:fz}
  f_{\rm x}(z')=\exp \left[
    3\int_0^{z'}\frac{1+w(\bar{z})}{1+\bar{z}}\mathrm{d}\bar{z}
  \right]\,.
\end{eqnarray}
For the concordance paradigm, namely the $\Lambda$CDM model, the DE equation of state is $w(z)\equiv-1$ corresponding to a cosmological constant $\Lambda$.
Thus, $f_{\rm x}\equiv1$ and $\Omega_{\rm x}\equiv\Omega_\Lambda$.
In the following, the choice $w(z)\equiv-1$ is adopted, unless otherwise specified.

The above isotropic energy output can be corrected for the beaming (see Sec.~\ref{progenitor}), once the jet opening angle $\theta$ is known, leading to beam corrected energy
\begin{equation}
E_\gamma = (1-\cos\theta) E_{\rm iso}\,.
\label{egdl2}
\end{equation}

It is important to stress that the prompt emission is not limited to the $\gamma$-rays and that, differently from the afterglow emission starting $\sim100$~s after the GRB trigger, current information in other energy bands is extremely difficult to observe without fast triggering.
Observations at lower energies (optical and X-rays) have been enabled only for GRBs with a precursor or a very long prompt emission duration, which gave the possibility of performing fast pointing to the source during the prompt phase \cite{Burlon2008}.

Regarding the GeV energy domain, a delayed (with respect to the trigger), long lived emission ($\gtrsim 10^2$~s), and separate lightcurve \cite{2019ApJ...878...52A} with a decaying luminosity as a power law in time, $L_{\rm GeV} \propto t^{-1.2}$ has been observed \cite{2019ApJ...878...52A}.
These distinctive features point towards a separate origin of the GeV with respect to the lower energy photons.

\begin{figure}[t]
\centering
\includegraphics[width=0.7\hsize]{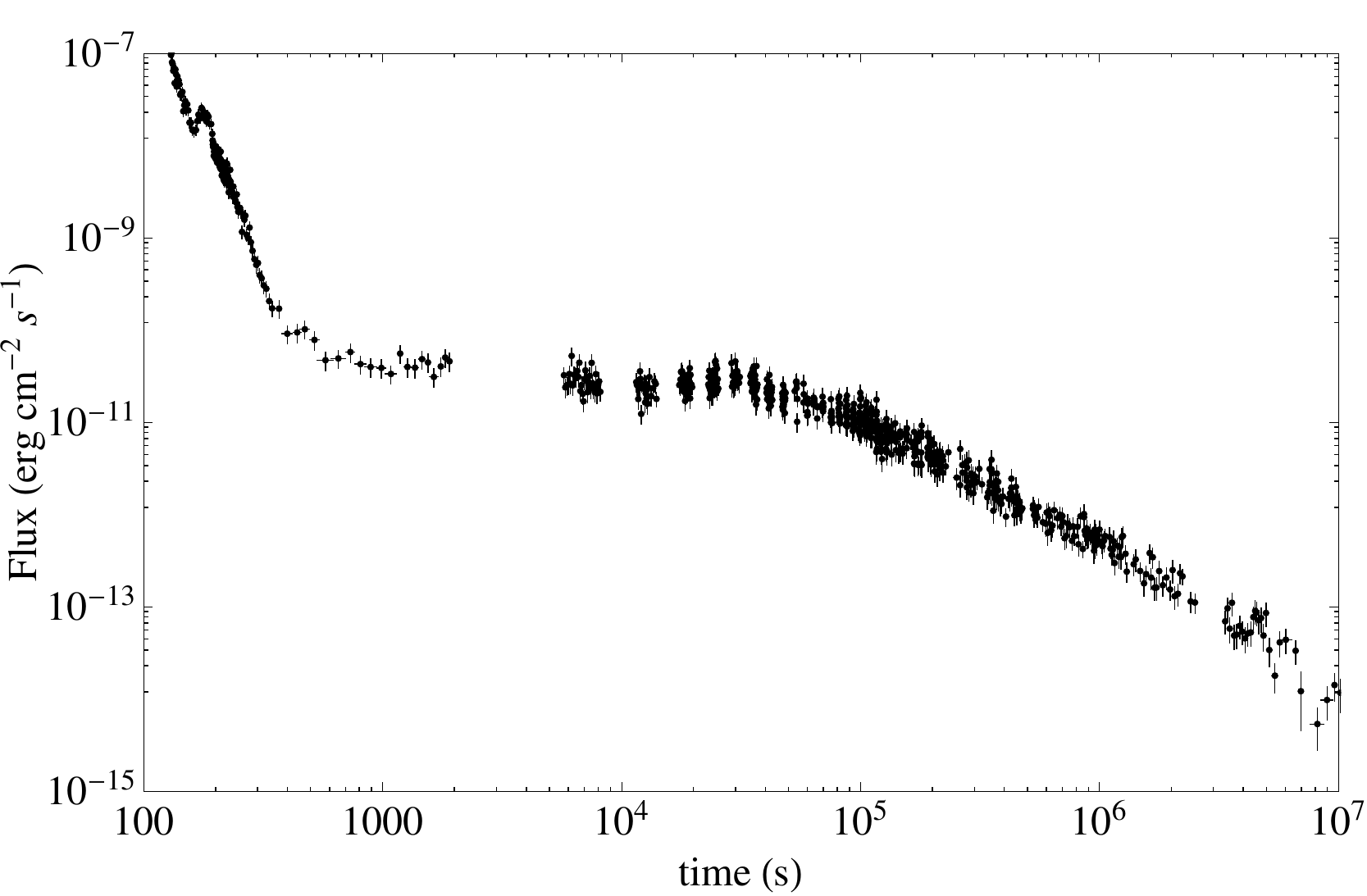}
\caption{\footnotesize The $X$-ray afterglow of GRB 060729 with all the three power-law segments and an initial flare clearly shown.}
\label{fig:Xlum}
\end{figure}

After $\sim100$~s since the trigger the prompt emission starts to decay in flux and, in many cases, this feature is caught by X-ray detectors \textit{Swift}-XRT within  $0.3$--$10$~keV energy band.
In general X-ray afterglow light curves show complex behaviors \cite{2012grb..book.....K} consisting of (see Fig.~\ref{fig:Xlum}):
\begin{itemize}
    \item[(1)] an early steep decay, interpreted as the tail of the prompt emission at large angles, followed by
    a very shallow decay, called the \textit{plateau}, usually accompanied by spectral parameter variations, and
    \item[(2)] a final decay, less steep than the first one.
\end{itemize}
The X-ray afterglow is also characterized by the presence of a flaring activity \citep{2012grb..book.....K}.
The observed behavior of these flares, the rapid rise and exponential decay together with a fluence comparable in some cases to the prompt emission, points out that the same mechanism for the prompt emission is responsible for the flaring activity \citep{2012grb..book.....K}.
Concluding, as above already stressed, these X-ray afterglow features are important to understand the nature of GRB progenitors.

\subsubsection{Timescales and characteristic energy as observable signature of GRBs}

There are other GRB observable quantities often employed in the literature, e.g., to construct GRB correlations (see Sec.~\ref{GRBcosmo}). They span from characteristic energies to timescales measured in several wavelengths. More specifically, a selection of them is summarized below:
\begin{itemize}
\item $t_{\rm b}$, the time at which the late X-ray afterglow power-law decline suddenly steepens due to the slowing down of the jet until the relativistic beaming angle roughly equals the jet opening angle $\theta$.
\item $\tau_{\rm lag}$, the time lag is computed as the difference of arrival times to the observer of the high energy photons ($100-300\,{\rm keV}$) and low energy photons ($25-50\,{\rm keV}$).\footnote{The time lag is historically  computed in these energy bands which are the BATSE energy channels 3 and 1, respectively.}
\item $t_{\rm X}$, the rest-frame time,  defined by a broken power-law fit of the X-ray afterglow light curve, at which a late power-law decay after the plateau phase is established.
\item $\tau$, the rest-frame time marking  the end of the plateau phase, defined from a fit of the X-ray afterglow with a smooth function given in Ref.~\cite{RuffiniMuccino2014}.
\item  $F_{\rm X}$ and $F_0$ are the observed X-ray fluxes respective to $t_{\rm X}$ and $\tau$, whereas the corresponding rest-frame $0.3$--$10$~keV luminosities $L_{\rm X}$ and $L_0$ are computes as follows\footnote{ As a convention, the X-ray luminosities are computed in a rest-frame energy band with similar extrema with respect to the observed one; with this prescription their expression are simple, as portrayed in Eq.~\eqref{simply}.
\begin{equation}
\label{simply}
L_{\rm X/0} = 4\pi d_{\rm L}^2 F_{\rm X/0} \frac{\int_{0.3/(1 + z)}^{10/(1 +
z)}{E N_{\rm E}^{\rm X/0}(E) dE}} {\int_{0.3}^{10}{E N_{\rm E}^{\rm X/0}(E) dE}} = 4\pi d_{\rm L}^2 F_{\rm X/0} (1+z)^{\gamma-2}\,,
\end{equation}
where we used the fact that X-ray data are observed by the \textit{Swift}-XRT in the $0.3$--$10$~keV energy band and the SED is in general a power-law spectrum with $N_{\rm E}^{\rm X/0}(E)\propto E^{-\gamma}$ and power-law index $\gamma>0$.}
\item $V$, the variability of the GRB light curve. It is computed by taking the difference between the observed light curve and its smoothed version, squaring this difference, summing these squared differences over time intervals, and appropriately normalizing the resulting sum.
\end{itemize}

\section{Theory of GRB progenitors}
\label{progenitor}

GRBs require progenitor systems able to guarantee enough energy for their powerful explosions to occur and emission mechanisms that can explain the above discussed spectral features. Although essential to better understand the physics of GRBs, neither a clear evidence for consolidate classes of suitable progenitors, nor a definitive GRB model have been yet established, as above stressed. However, observations, in the form of GRB spectra and light curves (see Sec.~\ref{sec:2.5}) and correlations between observable quantities (see Secs.~\ref{GRBcosmo} and \ref{sec:7}), enhanced our comprehension of these phenomena and  led to a general agreement on a few aspects below listed \cite{Piran2005}:
\begin{itemize}
\item[-] GRB progenitors harbor a BH\footnote{An alternative scenario proposes that a NS remnant could be left after a GRB emission, though this issue is still under debate. For details see Refs.~\cite{1992Natur.357..472U,Bucciantini2008,2012grb..book.....K} and Sec.~\ref{sec:magnetized_flow}.} which acts as a central engine powering the GRB emission.
\item[-] The burst energy must be gravitational and it is released in a very short time and from a compact region.
\item[-] Substantial part of this energy is converted into kinetic energy and a relativistic jetted outflow is formed.
\item[-] The acceleration process and the role played by magnetic fields are still unclear.
\item[-] The dissipation of part of the kinetic energy produces the observed prompt emission.
\item[-] The thermal emission of the prompt emission, may be the relic of the photons emitted during the initial explosion, whose energy has not been converted into kinetic form.
\item[-] Afterward, relativistic jets interact with the CBM, gradual energy conversion occurs and the afterglow (from X-ray down to radio) is produced.
\end{itemize}

The observed spectra have a considerable amount of $\gamma$-ray photons.
Photons with high energy $E_1$ annihilate with those at a low energy $E_2$ and produce $e^+e^-$ pairs if $\sqrt{E_1 E_2}\gtrsim m_ec^2$ (up to an angular factor), where $m_e$ is the electron mass.
If GRBs were not relativistic sources, the observed light curve variability time scale of $\delta t\approx10$~ms would imply that their emission would originate from a very compact region not larger than $R=c\delta t\approx3000$~km.
For typical values of the luminosity distance $d_L\approx 3 {\rm Gpc}\approx 10^{22}$~cm and fluence $S\approx 10^{-7}{\rm erg\,cm^{-2}}$ (energy at the detector per unit area) of GRBs, the opacity for pair creation is enourmous and it is given by \cite{Piran2005}
\begin{equation}
\label{photdens}
\tau_{\gamma\gamma}= f_{e^\pm}\frac{\sigma_T d_l^2 S}{m_e c^2 (c \delta t)^2}\approx10^{14}f_{e^\pm}\,,
\end{equation}
where $f_{e^\pm}$ is the fraction of photons with energies sufficient to produce pairs and $\sigma_T$ is the Thomson cross-section.
Such a large optical depth would imply that that the source must be optically thick leading to a thermal spectrum.
On the contrary, observations indicate that GRB spectra are typically non-thermal, pointing to the opposite conclusion that their source must be optically thin.
This issue is named \textit{compactness problem} \cite{Piran2005}.

However, the problem is only apparent, once relativistic effects are taken into account.
In fact, the causality limit of a source moving relativistically with bulk Lorentz factor $\Gamma\gg1$ towards the observer is $R\leq \Gamma^2c\delta t$.
Consequently, the observed photons are blue-shifted and their energy at the source is lower by a factor $\approx 1/\Gamma$, which may be insufficient for pair production. This leads to a decrease in the opacity, by a factor $\Gamma^{-2(\beta+1)}$, where the $\beta$ is the high-energy power-law index of photon spectrum of the burst.
For $\Gamma\gtrsim100$ one obtain the optically thin condition of the source.
Ultra-relativistic expansion of GRBs is unprecedented in astrophysics.
There are indications that relativistic jets in active galactic nuclei have $\Gamma\sim2$--$10$, but some GRBs have $\Gamma\gtrsim100$.
These large expansion velocities in GRB outflows find confirmations from the radio scintillation observed in their afterglows, and also from the apparent observation of self-absorption in the radio spectrum of the afterglow, where it is possible to obtain independent estimates of the dimensions of the afterglow relic \cite{2012grb..book.....K}.

\subsection{The fireball model}

The GRB standard model considers a homogeneous \textit{fireball}, \cite{Piran2005}.
For a pure radiation fireball, a large fraction of the initial energy released by the newly-formed BH is converted directly into photons.
Close to the BH, at a radius $r_0$ larger than the Schwarzschild radius, $R_S=2GM/c^2$, the photon temperature is
\begin{equation}
T_0 = \left(\frac{L}{4\pi\,a\,c\,r_0^2}\right)^{1/4} = 1.2\,L_{52}^{1/4} r_{0,7}^{-1/2}\,{\rm MeV}
\label{eq:T0}
\end{equation}
where $a$ is the radiation constant, and the luminosity $L$ and the radius $r_0$ are expressed, respectively, as $L_{52}=L/10^{52}$~erg/s and $r_{0,7}=r_0/10^7$~cm. In the following, to understand the order of magnitude of the key physical parameters characterizing GRBs, we use the notation $Q_x=Q/10^x$, where the quantity $Q$ is given in cgs units.
The temperature $T_0$ is above the threshold for pair production, hence a large number of $e^\pm$ pairs are created via photon-photon interactions, leading to a fully thermalized pairs-photons plasma with the opacity in Eq.~\eqref{photdens}.\footnote{Astrophysical fireballs include also some baryons from the surrounding medium, remnant of the progenitor system.}

GRB luminosities are many orders of magnitude above the Eddington luminosity, $L_E = 1.25 \times 10^{38} (M/M_\odot)
{\rm ~erg\, s^{-1}}$, therefore, the radiation pressure is much larger than self gravity and the fireball expands under its own pressure up to $\Gamma\approx10^2$--$10^3$ \cite{Goodman1986,Paczynski1986}.
Since the final kinetic energy cannot exceed the initial explosion energy $E_{\rm tot}$, the maximum attainable Lorentz factor is defined as $\Gamma_{\rm max} = E_{\rm tot}/M c^2$ and depends upon the amount of baryons (baryon load) of rest mass $M$ within
the fireball \cite{Piran2005}.

\subsubsection{Photon-dominated scenario}

The simplest scenario considers a photon-dominated expanding shell of width $\delta r^\prime$ ``instantaneously'' releasing its energy.
From here on, prime symbols indicate quantities measured in the comoving frame of the shell, namely from an observer within it.
On the other hand, $r$ is the radial coordinate of the laboratory frame, a frame outside the shell where the observer is sitting on the central engine.

Enforcing energy and entropy conservation laws, the shell keeps accelerating up to $\Gamma_{\rm max}\simeq \eta$, which is attained at at the so-called dissipation radius $r_s \sim \eta r_0$; beyond it, most of the internal energy of the shell has been converted into the kinetic one, so the flow no longer accelerates and it coasts.
Thus, the fireball obeys the following  scaling laws of the shell comoving temperature, Lorentz factor and comoving volume, respectively
\begin{equation}
\left\{
\begin{array}{lllll}
T^\prime(r) \propto r^{-1} \quad &, \quad \Gamma(r) \propto r \quad &, \quad V^\prime(r) \propto r^3\quad   & , \quad & r<r_s\\
 T^\prime(r) \propto r^{-2/3} \quad &, \quad \Gamma(r) = \eta \quad &, \quad V^\prime(r) \propto r^2 \quad   & , \quad & r\gtrsim r_s
\end{array}
\right.\,.
\end{equation}
from which follows that as the shell accelerates (as $\Gamma$ increases with $r$), its internal energy drops (as $T^\prime$ decreases with $r$).
Finally, the evolution the observed temperature is given by
\begin{equation}
T^{ob}(r) = \Gamma(r) T^\prime(r) =
\left\{
\begin{array}{lll}
T_0                 \quad   &,  \quad   & r< r_s \\
T_0 (r/r_s)^{-2/3}  \quad   &,  \quad   & r\gtrsim r_s
\end{array}
\right.\,.
\label{eq:8}
\end{equation}

\subsubsection{Internal shock scenario and photospheric emission}
\label{sec:scalings2}

In the case of LGRBs, the progenitor continuously emits energy at a rate $L$, over a longer duration $t \gg r_0/c$, and ejects mass at a rate $\dot{M} = L/(\eta c^2)$.
In this case, the scaling laws for the instantaneous release are still valid, provided that $E$ is replaced by $L$ and $M$ by $\dot M$, and a further equation for the mass conservation of the baryons (within the spherical symmetry assumption) is required \cite{Peer2010}
\begin{equation}
n^\prime_p(r) = \frac{\dot{M}}{4\pi r^2 m_p c \Gamma(r)} = \frac{L}{4 \pi r^2 m_p c^3 \eta \Gamma(r)}\,,
\label{eq:n}
\end{equation}
where $n^\prime_p(r)$ is the comoving number density of baryons and $m_p$ is the proton mass.

For longer activity of the inner engine, fluctuations in the energy emission rate would result in the propagation of independent shells, each of them with analogous thickness $r_0$ and dynamics.
For two consecutive shells with a difference in their Lorentz factors $\delta \Gamma \sim \eta$ or velocities $\delta v \approx c/(2 \eta^2)$, collisions become possible after a typical time $t_{col} = r_0/\delta v$ and an observer frame radius \cite{1993ApJ...415..181M}
\begin{equation}
r_{col} = v t_{col} \simeq c t_{col} \simeq 2 \eta^2 r_0\,.
\end{equation}
Above $r_{col}$, which is a factor $\eta$ larger than $r_s$, collisions occur, dissipate the kinetic energy and convert it into the observed radiation
\cite{ReesMeszaros1992,SariPiran1997}.
The advantages of the internal shock scenario are listed below.
\begin{itemize}
\item[1.] {\bf Light curve variability.} The time delay between the photons produced by the collisions and a photon emitted from the central engine towards the observer, i.e., $\delta t^{ob} \simeq r_{col}/(2 \eta^2 c) \sim r_0/c$,
is similar to the central engine variability and can explain the observed variability ($\gtrsim 1$~ms).
\item[2.] {\bf Particle acceleration.} Shell collisions generate internal shock waves, which can accelerate particles to high energies via Fermi mechanism and produce $\gamma$-rays.
\item[3.] {\bf Thermal radiation.}  Eq.~\eqref{eq:T0} states the fireball is optically thick \cite{Paczynski1986,Goodman1986,ShemiPiran1990}.
For $r>r_s$, an effective photosphere radius $r_{ph} \simeq L\sigma_T/(8 \pi m_p c^3\Gamma\eta^2) \simeq 2 \times 10^{11}~L_{52}\,\eta_{2.5}^{-3}~{\rm cm}$ can defined by requesting $\tau(r_{ph}) = 1$ \cite{Paczynski1991}.
Internal shocks take place at $r_{col} \sim \eta r_s > r_{ph}$.
In a more realistic picture, photons decouple the plasma on ``photospheric surface'' \cite{Peer2008} and the emerging emission is a convolution of different Doppler boosts and different adiabatic energy losses of photons \cite{Peer2008,Ryde2011}. This emission explains the thermal-like emission embedded in the non-thermal spectra of some GRBs \cite{Ryde2004,Ryde2009,Ryde2011}.
\end{itemize}
However, the internal shock scenario manifests some drawbacks.
\begin{itemize}
\item[1.] {\bf Efficiency.} From the energy and momentum conservation, the kinetic energy dissipation is highly efficient only if two shells have masses $m_1 \simeq m_2$ and Lorentz factors $\Gamma_1/\Gamma_2\gg 1$. The average over several collisions leads to a low global efficiency of $1$--$10\%$ \cite{Kobayashi1997,KobaSari2001}, which contrasts to the much higher efficiency $\sim 50\%$ inferred from afterglow measurements \cite{Nousek2006,Zhang2006}. Higher efficiency up to the $\sim60\%$ can be attained by considering larger contrasts $\Gamma_1/\Gamma_2\gg10$ \cite{KobaSari2001}. However, these Lorentz factor contrasts unlikely occur within the traditional collapsar or the merger scenarios.
\item[2.] {\bf Observed spectra.} This model does not explain the observed spectra and needs further assumptions on how the dissipated
energy produces photons (i.e., involving standard radiative processes such as synchrotron emission or Compton scattering).
\end{itemize}

\subsubsection{Magnetized (or Poynting-flux dominated) outflows}
\label{sec:magnetized_flow}

The Poynting-flux dominated model speculates that the gravitational energy produces very strong magnetic fields, which may be crucial in the jet formation of GRBs, similarly to the Active Galactic Nuclei (AGN), where magnetic energy is converted into particle acceleration via Blandford-Znajek \cite{1977MNRAS.179..433B} or Blandford-Payne \cite{1982MNRAS.199..883B} mechanisms.
The idea behind this model is that the collapse of a white dwarf (WD) induced by accretion from a massive star, the core collapse of a massive star, or NS merger does not form immediately a BH, but rather a rapidly-spinning (with a period of $\sim 1$~ms) and highly-magnetised NS (with a magnetic field $B \gtrsim 10^{15}$~G) NS, known as \textit{magnetar} \cite{1992Natur.357..472U}.
The maximum amount of magnetic energy that can be stored is $\sim 2 \times 10^{52}$~erg and it can be extracted in a short timescale of $\sim 10$~s and drives a
jet along the polar axis of the NS powering the prompt emission \cite{Bucciantini2008}.
The decay of rotational or magnetic energy may continue
to power late time flaring or afterglow emission.
The dipole radiation naturally produces a plateau phase up to the dipole spin-down time scale \cite{2012grb..book.....K}.

In this model, the magnetic field is essentially toroidal (i.e., $\vec B \perp \vec \beta$) and its polarity in the flow changes on small scale defined by the light cylinder in the central engine.
The total luminosity is given by $L = L_k + L_M$, where $L_k = \Gamma \dot{M} c^2$ is the kinetic part and $L_M=4\pi r^2 c [B^2/(4\pi)]$ is the magnetic part \cite{Giannios2005,Giannios2006}. The key parameter is the magnetization $\sigma \equiv L_M/L_k = B^2/(4 \pi \Gamma^2 n m_p c^2)$, which plays a similar role to the baryon load in the classical model and defines the maximum attainable Lorentz factor $\Gamma_{\max} \approx \sigma^{3/2}$, whereas, during the acceleration phase, one gets $\Gamma(r) \propto r^{1/3}$ \cite{Giannios2005,Giannios2006}.

In this model the rapid variability observed in GRBs and the low efficiency in dissipating the kinetic energy via shock waves in highly magnetized plasmas are still open issues.
Recent recipes suggest that central engine variability leads to the ejection of magnetized plasma shells which expand due to internal magnetic pressure gradient and collide at a distance $r_{col}$. The ordered magnetic field lines of the ejecta get distorted and fast reconnection occurs. The induced relativistic turbulence may be able to overcome the low efficiency difficulty of the classical internal shock scenario \cite{2011ApJ...726...90Z}.

\subsubsection{Particle acceleration}
\label{sec:acc}

To produce the non-thermal GRB spectra, part of the kinetic energy needs to be dissipated and used to increase the random motion of the outflow particles and/or accelerates some fraction of them to a non-thermal distribution.
Once accelerated, these high-energy particles emit non-thermal photons.

The most widely proposed particle acceleration mechanism within the internal shock scenario is the Fermi mechanism \cite{Fermi1949}.
In this process, the accelerated
particles cross the shock fronts, and during each crossing their energy increases at a constant rate $\Delta E / E \sim 1$. The accelerated particles have a power-law energy distribution $N(E) \propto E^{-\delta}$ with index $\delta\approx 2.0$--$2.4$ \cite{EllisonDouble2004}.

Dissipation mechanisms in magnetized outflows have been longly discussed (see Ref. \cite{2015AdAst2015E..22P} and references therein).
%
%
%
%
%
Further, particles may also be accelerated via Fermi mechanism in shock waves, but it has been pointed out that in highly
magnetized plasma this process may be inefficient \cite{2011ApJ...726...75S}.

\subsubsection{Radiative processes}
\label{sec:radiation}

After kinetic energy dissipation and particle acceleration, energy conversion is needed to produce the non-thermal spectra observed in GRBs.
The most discussed radiative model in the literature is the optically thin synchrotron emission \cite{ReesMeszaros1992,MeszarosRees1993,Tavani1996,Sari1997,DaigneMochkovitch1998}, accompanied by synchrotron-self Compton (SSC) at high energies \cite{Chiang1999,Stern2004,2009ApJ...703..675N}.

Recently, with mounting evidences of thermal components in GRB spectra \cite{Ryde2005,Guiriec2011}, the photospheric model has acquired growing relevance \cite{MeszarosRees2000,Peer2006,Ryde2010}.
This model is not in constrast with and has to be considered complementary to the synchrotron+SSC emission, which originates from a different region of the outflow.
\\

{\bf Synchrotron emission}

The synchrotron emission has been extensively studied for interpreting first the non-thermal emission in AGNs and then the GRB afterglow emission \cite{RybickiLightman}. Regarding the explanation of the GRB prompt emission spectra \cite{ReesMeszaros1992,MeszarosRees1993b,Tavani1996,Sari1997,DaigneMochkovitch1998}, the synchrotron emission model has several advantages:
\begin{itemize}
\item[1)] requires energetic particles and strong magnetic fields, both expected in shock waves;
\item[2)] has a broad-band spectrum with characteristic peak, associated with the observed peak energy;
\item[3)]  for typical parameters, energetic electrons radiate nearly 100\% of their energy.
\end{itemize}

A source at redshift $z$, expanding with $\Gamma=\sqrt{1-\beta^2}$ and at an angle $\theta$ with respect to the observer, emits photons which are seen with a Doppler boost $\mathcal{D} =[\Gamma(1-\beta\cos\theta)]^{-1}$.
In the comoving frame, electrons move in a magnetic field $B$ and, thus, have random Lorentz factor $\gamma_{e}$. Their typical energy is \cite{RybickiLightman}
\begin{equation}
\varepsilon_{ob} = \frac{3 q \hbar B\gamma_{e}^2}{2m_e c}  \frac{\mathcal{D}}{(1 +z)}= 1.75\times 10^{-19} B \gamma_{e}^2 \frac{\mathcal{D}}{(1+ z)}\,{\rm erg}\,.
\label{eq:nu_m}
\end{equation}
Typical GRB peak energies $\varepsilon_{ob}\approx 200$~keV require strong magnetic fields and very energetic electrons, both feasible for Poynting flux-dominated outflows or photon-dominated outflows where strong magnetic fields may be generated via Weibel instabilities \cite{Spitkovsky2008}.\footnote{However, both $B$ and $\gamma_{e}$ are much higher than the ones inferred from the fit with the synchrotron model of the GRB afterglow, whose microphysics of particle acceleration and magnetic field generation should be similar to that of the prompt emission environment \cite{KumarPanaitescu}.}

On the other hand, strong magnetic fields imply the comoving cooling time of the electrons to be $t^\prime_{cool}\lesssim t^\prime_d \sim R/(\Gamma c)$. Thus the expected synchrotron spectrum below the peak energy would be $F_\nu\propto\nu^{-1/2}$ (or $N_E \propto E^{-3/2}$) \cite{Sari1998,Sari1998b}, which is inconsistent with the average low energy spectral slope $\langle\alpha\rangle=-1$ (see Fig. \ref{figBand}) and, hence, the value $\alpha=-3/2$ is called ``synchrotron line of death''.
To overcome this problem, electrons must cool slowly, leading to a spectrum below the peak given by $F_\nu\propto\nu^{1/3}$ (or $N_E \propto E^{-2/3}$), which is roughly consistent with the observations.
However, the condition $t^\prime_{cool}\gtrsim t^\prime_d$ leads to high values of $\gamma_e$, whereas $B$ would be very low and, in order to explain the observed flux, the electron energy would be several orders of magnitude higher than that stored in the magnetic field \cite{Kumar2008}. To overcome this, the inverse Compton contribution has to be significant, producing $\sim$~TeV emission. To avoid a substantial increase of the total energy budget, the emission radius should be $R \gtrsim 10^{17}$~cm, but cannot explain the rapid variability observed \cite{Kumar2008}. Suggested modifications (and drawbacks) to the synchrotron scenario can be found in the literature \cite{Granotetal2000,Panaitescu2000,Peer2006b,Piran2009,2009ApJ...703..675N,Gupta,2012ApJ...757..115A}.\\
%
%
%
%
%
%
%

{\bf Photospheric emission}

For $r_{ph} \gg r_s$, a large fraction of the kinetic energy is dissipated below the photosphere \cite{2004ApJ...613..448P}. The produced non-thermal photons cannot directly escape and are advected with the flow until the transparency.
Within the flow, multiple Compton scatterings occur and modify the synchrotron spectrum of the heated electrons, which rapidly cool, mainly by IC scattering. The electron distribution becomes quasi-Maxwellian, with a temperature determined by the balance between heating (external and by direct Compton scattering of energetic photons), and cooling (adiabatic and radiative) \cite{Peer2005}.
Finally, the photon field is modified by the scattering from the quasi-Maxwellian
electron distribution \cite{2004ApJ...613..448P}.

Further, the thermal photons of the fireball contribute as seed for IC scattering, hence the non-thermal electrons, heated by energy dissipation below the photosphere, rapidly cool and reach a quasi-steady state distribution \cite{Peer2006}. The result is a \textit{two temperature plasma}, with electron temperature $T_e > T_{ph}$.
If dissipation processes occur at intermediate optical depth $\tau\sim$few--few tens, the resulting spectrum is:
\begin{itemize}
\item[1)] similar to the Rayleigh-Jeans part of the thermal spectrum, for $T<T_{ph}$;
\item[2)] $F_\nu \propto \nu^{-1}$
(or $N_E \propto E^{-2}$) because of
multiple Compton scattering, for $T_{ph}<T<T_e$;
\item[3)] an exponential cutoff, for for $T>T_e$.
\end{itemize}
The spectral slope obtained in the above 2) is similar to the high energy spectral slope in GRB spectra, $\langle\beta\rangle \sim -2$, thus, it could be concluded that $E_{p}$ is associated with $T_{ph}$.
However, recently Fermi data has shown thermal peaks at lower energies than $E_{p}$, which points rather to the more natural interpretation that the thermal peak is associated with $T_{ph}$ and suggests that $E_{p}$ may be associated with $T_{e}$ or with the synchrotron emission.
Moreover, if dissipation occurs at $\tau \gtrsim 10^2$, the resulting spectra is thermal-like.
On the other hand, for $\tau \lesssim$~a few a more complex spectrum forms, with the main contribution coming from synchrotron photons (emitted by the electrons) below the thermal peak and above it from multiple IC scatterings (leading to a nearly flat energy spectrum) \cite{Peer2005}.
All the above discussions is viable for dissipation processes from highly magnetized plasma as well \cite{Giannios2006,2015ApJ...801..103G}.

However, also the above model suffers two major drawbacks, since it cannot explain
\begin{itemize}
\item[1)] low energy spectral slopes less steep than the Rayleigh-Jeans part of a Planck spectrum;
\item[2)] the observed GeV emission, which may originate from some dissipation above the photosphere.
\end{itemize}

\section{Reconciling cosmological indicators to GRBs}\label{sec:3}

After the first part of this review, in which we faced the main properties of GRBs,  their possible theoretical background and progenitors, we are now in condition to relate our understanding directly to cosmology. In fact, for cosmological purposes it is essential to get the distance of astronomical objects and so the use of GRBs would help in computing such distances up to very large redshifts. In particular, source physical parameters mostly depend on  luminosity and size and then cosmic bounds can be inferred if there exists a relation between distances and redshifts. This prerogative is intimately related to two distinct concepts, i.e. distance indicators\footnote{Examples are main sequence fitting method, variable stars, Tully-Fisher and Faber-Jackson relations, etc.} and standard candles\footnote{Examples are SNe Ia, Cosmic Microwave Background (CMB) measurements, Baryonic Acoustic Oscillations (BAO), etc.}.

Below we elucidate the main properties of such objects and the most important consequences they have in observational cosmology.

\subsection{Distance indicators}

At the beginning of our review, we emphasized how distances in cosmology are relevant to compute GRB luminosity/energy. A further step consists in noticing the distances measurements are classifiable by
\begin{itemize}
\item[\,] {\bf Absolute measures}, as they are computed through previously known information, e.g. trigonometric parallax.
\item[\,] {\bf Relative measures}, as they involve empirical relations based on indirect or direct probes, e.g. Cepheids period-luminosity relation, for which the distance measures are calibrated against an absolute method to enable those measurements to be somehow \emph{anchored}.
\end{itemize}

Standard cosmology shows up how to relate the redshift to metric distances in both the above cases.
The machinery of  dynamical distance indicators involves tightly packed all the
ingredients of cosmological physics. We thus require the cosmological principle to hold in an expanding universe in the context of general relativity. Despite obvious, there is no direct analogy to classical dynamical
distance indicators, as the laboratory in which measurements are got is moving as well. Precision cosmology would enrich data during incoming years, as future surveys will provide  resource of
data to constrain and refine our understanding about distances and cosmological parameters.

Using current data catalogs, it appears evident that GRBs can be significantly investigated once the calibration of the correlation functions are
deduced from  absolute confidence. Recently, techniques of non-calibration have been more often used, overcoming the problem of  standardizing GRBs that are, as known, not perfect standard candles for cosmological distance tests. Later on we confront the calibration and uncalibration procedures, emphasizing how to single out the most promising treatment to handle GRBs in cosmology.

\subsection{Standard candles}

Above we stated astrophysical distances are crucial for picturing current universe. Though essential, estimating cosmic distances mainly remains a complicate prerogative. In view of the above classification, the distance estimation passes through the use of {\em standard candles}. These objects hold the fundamental property of relating the intrinsic luminosity, namely $L$, to some known property, enabling one to get constraints over it. Once the luminosity is known, the distance can be computed accordingly.

A standard procedure is to get measures of the energy emitted from astrophysical objects. The energy bounds are got in a precise time interval, say $\Delta t$ and by virtue of $E=L\cdot\Delta t$, {\it i.e.}, the relation between luminosity and energy, it is possible to get distances from the energy itself, through a well-consolidate strategy, below reported.

Detectors are able to catch fractions $E_d$ of the emitted energy $E$, which is proportional to the ratio between the detector area $A$ and the spherical shell $4\pi d_L^2$ in which one defines the cosmic distance $d_L$, {\it i.e.},
\begin{equation}
   E_{\rm d} = \frac{E\,A}{4\pi d_L^2}\,.
\end{equation}

A general relation for $d_L(z)$ is written as
\begin{equation} \label{dL SN}
 d_L (z, \theta) = c\,(1+z) \int_0^z \frac{\mathrm{d} z'}{H(z', \theta)} \,
\end{equation}
where the set of free parameters to constrain  is indicated by $\theta$. Exploring a given  cosmological model is equivalent to get $\theta$.

Thereby, combining the aforementioned quantities, we obtain the energy per unit detector area $A$ and per unitary time $\Delta t$, which defines the flux expressed by
\begin{equation}
F = \frac{E_{\rm d}}{A\,\Delta t} = \frac{L}{4\pi d_L^2}\,.
\label{InverseSquare}
\end{equation}
As we highlighted, the luminosity $L$ is known for standard candles, thus one can measure $F$ in order to get a given astrophysical object distance.

\subsection{Classifying standard candles}

We above stressed that physical laws underlying a particular astronomical object permit one to know the luminosity of standard candles.
Clearly, such rules are essentially based on thermodynamic or chemical processes of a given astrophysical object. Consequently, one can classify standard candles by means of these physical laws and, according to the simplest classification scheme, we can handle at least two kinds of standard candles summarized below \cite{2008cosm.book.....W}.
\begin{itemize}
\item {\em Standard candles as primary distance indicators}, which can be calibrated within the Milky Way galaxy.
\item {\em Standard candles as secondary distance indicators}, which can be observed at larger distances than Milky Way scales. However, they require calibration, typically performed  using known primary distance indicators within distant galaxies.
\end{itemize}

\subsubsection{Primary distance indicators}

The above first typology mainly includes {\em Variable stars}, {\it i.e.}, among which, Cepheids, RR Lyrae and Mira. Here, the variable star type is based on the possible correlation  between their period of variation, steadily measured, and their luminosity. Even though this set of stars mainly constitutes the primary indicators, further typologies are main sequence and red clump stars. Here, using the luminosity-temperature relations from the standard  Hertzsprung-Russell diagram, one deduces stellar luminosity within a fairly narrow range. Last but not least, eclipsing binaries are also primary distance indicators, since their  luminosity is computed by the Stefan-Boltzmann law through a direct estimate of their radius, by means of a Doppler measurement of orbital velocities combined with the light-curve data, together with the temperature, deduced from the spectrum.

\subsubsection{Secondary distance indicators}

On the other hand, the second class of standard candles are essentially based on very different indicators with respect to the first case. For instance, the prototypes of such indicators are the {\em properties of galaxies}, among all, the Tully-Fisher relation. This law matches spiral galaxy rotation speed and stellar luminosity. In particular, to argue the spiral galaxy rotaion speed one can consider, for example, the spectral line width. Another relation, widely adopted as underlying second type of indicators, is the Faber-Jackson relation. Here, it is possible to infer  elliptical galaxy random stellar velocities using the total luminosity. Again, the way to get these velocities consists in the use of spectral line widths. Another quite relevant relation is  the fundamental plane law, {\it i.e.}, a treatment that extends the Faber-Jackson one by including surface brightness as an additional observable parameter.

Besides galaxy properties, another second typology of standard candles is represented by  {\em SNe Ia}, {\it i.e.}, probably the most used cosmological standard candles to accredit the late time cosmic speed up. The scenario in which they form is due to thermonuclear explosions of WDs that exceed the Chandrasekhar's limit, namely $\sim1.4 M_{\odot}$.
For such objects, we see a correlation between the time scale of the explosion and the peak luminosity.
The corresponding light curves follow given shapes, in agreement with the so-called Phillips curve \cite{Phillips1993}.
As stated, SNe Ia are the most fruitful standard candles. For each event, even if the luminosity is clearly  different for every SN, the Phillips curve relates the B magnitude peak to the luminous decay after $15$ days with an overall set of SNe distributed in the range $z=0$--$2.5$. These redshifts span  between decelerating and accelerating phases of universe's evolution, corresponding to the matter and DE dominated epochs\footnote{They refer to ``Type Ia'' for the absence of hydrogen and the presence of once ionized silicon (SiII) in their early-time spectra.}. Last but not least, these indicators are present in all galaxies, except in the arms of spiral galaxies, but their physical internal processes are still object of investigations as they are not fully-interpreted.

\section{Going ahead with standard indicators: the $\chi^2$ analysis}

Using standard candles, it is possible to establish data catalogs that can be used and matched with GRB data. Hence, to experimentally fit a given model with a given set of free parameters, one requires the definition of a merit function that  quantify the overall agreement between the working model with the aforementioned cosmic data.  Equivalently, it is of utmost importance to get best fit parameters  and corresponding estimates of error bars, together with a method to possibly measure the goodness of fit. The parameter fitting treatment commonly makes use of  least-squares analyses, based on the combination among  data points, say $D_i$, a model for these data, namely $y(x,\vec{\theta})$, function of $\theta$. Naively the simplest approach to least squares for uncorrelated data becomes
\begin{equation}
\chi^2=\sum_i w_i [D_i-y(x_i|\vec{\theta})]^2
\end{equation}
where the weights $w_i$ reach the maximum variance in case $w_i=1/\sigma_i^2$,  with $\sigma_i$ the data point errors. For correlated data, we have
\begin{equation}
\chi^2=\sum_{ij} (D_i-y(x_i|\theta))Q_{ij}^{-1}(D_j-y(x_j|\theta))
\end{equation}
in which the inverse of covariance matrix, $Q$, has been introduced describing the degree of correlations among data. Minimizing the $\chi^2$ is equivalent to get suitable sets of findings that represent the best fit for our procedure. Different $\chi^2$ values lead to probability distribution around the minimum.

\subsection{Probability distribution}

Analyzing the probability distribution, once the above treatment is worked out becomes essential. In particular, probabilities $p$ that the observed $\chi^2$ exceeds by chance a value $\widehat{\chi}$ for the correct model  is clearly calculable and, in fact, $Q$ provides a measure of the goodness of fit, as one infers it at the minimum of $\chi^2$. Two limiting cases, unfortunately, are possible, $Q$ is too small or too large. The first occurrence leads to the fact that the model is either wrong or errors are  underestimated and/or they do not distribute Gaussianly. The second occurrence happens when either errors are overestimated or data are correlated while rarely it could also happen the distribution is non-Gaussian.

In general, the statistical procedure suggests $\chi^2$ roughly comparable with the data number. Consequently, using the reduced chi square, as the ratio between the chi square and the number of degrees of freedom, could be an useful trick to handle experimental workarounds.

\subsection{The SNe Ia measurements}

SNe are widely-adopted in astrophysics as standard candles. Thereby, several SN catalogs are often updated, furnishing today a large number of data points that combined with other data sets enable one to fix tighter constraints over the universe expansion history in terms of its constituents.  In particular, SNe Ia are likely the most used objects that constrain DE at late times. The standard procedure makes use of the luminosity distance $d_L(z)$ and of apparent magnitude.  A general relation for $d_L(z)$ has been previously written, with $\theta$ the set of free parameters of a given model. Then, we can notice that exploring a given  cosmological model is equivalent to get the whole set of parameters, $\theta$.

In particular, when one adopts a given cosmological model then  an indirect requirement  naturally holds: \emph{the underlying  cosmological model is the most suitable one}. This is clearly a limitation because this hypothesis do not always coincide with the most feasible statistical model. So, more than one scenario can lead to subtle bounds, indicating a degeneracy problem among different models. This justifies the need of analyzing different cosmological paradigms working out data set hierarchy, {\it i.e.}, combining more than one data catalog. In addition, statistical criteria are also crucial to check the goodness of a given paradigm.

For SNe Ia, by virtue of Eq.~\eqref{dL SN}, it is possible to relate the brightness to fluxes to get the distance modulus
\begin{equation}
\label{defmu}
 \mu(z) = 25 + 5 \log \left(\frac{d_{\rm L}}{{\rm Mpc}}\right)\,.
\end{equation}
Neglecting  error bars on $z$, we underline errors on $\mu$, namely $\sigma_\mu$, whereas the  best fit is determined by the standard maximization of the underlying likelihood function, or simply minimizing the $\chi^2$, provided by
\begin{equation}\label{chisne1}
\chi^2 (\theta_{min}) = \sum_{i=1}^{N_s} \left[\frac{\mu_i(z_i, \theta) - \mu_{obs,i}(z_i)}{\sigma_{\mu,i}}\right]^2\,,
\end{equation}
where the subscript $min$ refers to the set of values that minimize the chi square function, as above requested. Theoretical models can be therefore tested by $\chi^2$ statistics, leading to probe DE by inferring $d_{L}$ in units of megaparsecs and using it by means of the apparent magnitude.

Again, intertwined more than one data set with other surveys is quite essential to determine the whole set of parameters, with refined accuracy. For instance, SNe alone, as well as GRBs\footnote{Weak similarities between GRBs and SNe Ia may occur at the level of formal computation, although the GRB nature is absolutely different from SNe. The core idea is to write a GRB luminosity distance as well  and proceed analogously.}, $H_0$ cannot be arguable. In fact, expanding up to the first order the luminosity distance, valid up to $z\lesssim 0.001$, one gets
\begin{equation}
d_{\rm L}(z,H_0)\simeq \frac{cz}{H_0}\,,
\end{equation}
that clearly vanishes at $z=0$, implying $H_0$ cannot be constrained with SNe Ia alone. Also,  a multiplicative degeneracy between $H_0$ and the other free parameters occurs.

Once computed the chi square statistic, the confidence regions are planes with fixed $\chi^2$. For example, one can get
$\Omega_{M}-\theta_i$ planes by  marginalizing
the likelihood functions over $H_{0}$. This procedure consists in integrating the
probability density $p\propto\exp(-\chi^2/2)$ for all values of
$H_{0}$.
Marginalization is a generic technique, clearly not limited to $H_0$. In fact, as one desires to simultaneously constrain a few parameters and meanwhile wants to get the corresponding probability distribution  regardless of the values of a given parameter, say $\theta^\star$, can proceed with marginalizing. Let us call  $\theta^\star$ the parameter we do not care about, the marginalized probability density, computed for example for $\Omega_m$, is given by
$p(\Omega_m)=\int d\theta^\star \,p(\Omega_m, \theta^\star)$.

\subsection{BAO measurements}

The BAO measurements are due to overdensity of baryonic matter due to acoustic
waves. These waves propagate in the early universe
\citep{Silk68,Peebles70} and represent \emph{standard ruler} for cosmological
length scale. This signature, in the large-scale clustering of
galaxies, constrains cosmological parameters by
detection of a peak in the correlation function
\citep{Eisenstein05}, by defining the $A$ parameter as follows
\begin{equation}
\label{ABAO}
A = \frac{\sqrt{\Omega_{m}}}{z_1}
 \left[\frac{z_1}{E({\bf x},z_1)}\frac{1}{|\Omega_k|} {\rm sinn}^2
 \left(\sqrt{|\Omega_k|}\int_0^{z_1}\frac{dz}{E({\bf x},z)}\right)\right]^{{1\over3}}\,,
\end{equation}
where ${\bf x}$ is the set of cosmological density parameters, $E({\bf x},z)=H({\bf x},z)/H_0$ and ${\rm sinn}(x)=\sinh(x)$ for the curvature parameter $\Omega_k>0$, ${\rm sinn}(x)=x$ for $\Omega_k = 0$, and ${\rm sinn}(x)=\sin(x)$ for $\Omega_k < 0$.
The $A$ parameter has been measured from the SDSS data and reads to be $A=0.469(0.95/0.98)^{-0.35}\pm
0.017$, with $z_1 = 0.35$, so the $\chi^2$ in terms of $A$ reads $
\chi^{2}_{\rm BAO}=(A-0.469)^2/0.017^2$.
The BAO corresponding angular distance measures can be defined by means of
\begin{equation}
\label{eq:DV}
d_{\rm z}({\bf x},z) \equiv r_{\rm s} (z_\text{d}) \left[\frac{c\,z}{H({\bf x},z)}\right]^{-1/3}\left[\frac{d_{\rm L}({\bf x},z)}{1+z}\right]^{-2/3}\,.
\end{equation}
%
%
The corresponding $\chi^2$ is given by
\begin{equation}
\label{chiBAO}
\chi^2_{\rm BAO}=\sum_{i=1}^{N_{\rm BAO}} \left[\frac{d_{\rm z}^{\rm th}({\bf x},z_i)-d_{\rm z,i}^{\rm obs}}{\sigma_{d_{\rm z,i}}}\right]^2\ .
\end{equation}

It is clear that BAO measures are slightly model-dependent as they depend on the comoving sound horizon $r_{\rm s}(z_\text{d})$. In particular, in Eq.~\eqref{eq:DV} the sound horizon depends upon the baryon drag redshift $z_\text{d}$. This quantity requires  calibration that typically is performed with CMB data, adopting a given  background model, that commonly is the $\Lambda$CDM scenario. Very often the best expected values are given by   $z_\text{d}=1059.62\pm0.31$ and $r_{\rm s}(z_\text{d})=147.41\pm0.30$ \cite{Planck2018}.

\subsection{Differential age and Hubble measurements}

Another intriguing treatment, widely used in observational cosmology and also for   calibrating GRB correlations, has been firstly proposed in  Ref.~\cite{2019MNRAS.486L..46A}. The idea is to measure the Hubble rate by using galaxies, in a quite model-independent way. In the context of GRBs, the Hubble catalog has been widely explored. For example, in Ref.~\cite{2019MNRAS.486L..46A}, the core idea is to match the observational Hubble rate data (OHD) with model independent expansion of $H$ made by B\'ezier polynomials.
At a first glance, this \textit{differential age method}  (see, e.g., Refs.~\cite{2002ApJ...573...37J,2016JCAP...05..014M}) does not require any assumption over the form of $H$, although spatial curvature can affect the overall treatment if it varies with time, instead of being fixed\footnote{We here focus on vanishing spatial curvature, {\it i.e.}, $\Omega_k=0$ \cite{Planck2018}.}.

To better introduce the method, we notice it is well-known that spectroscopic measurements of the age difference $\Delta t$ and redshift difference $\Delta z$ of couples of passively evolving galaxies lead to $    \Delta z/\Delta t\equiv dz/dt$ and so, if galaxies formed at the same time (redshift $z$), the Hubble rate can be approximated by
\begin{equation}\label{ehsi}
    H(z)=-\left(1+z\right)^{-1}\Delta z/\Delta t\,.
\end{equation}
Consequently, model-independent estimates may come from cosmic chronometers based on the assumption that observable Hubble rates are given by the exact formula
\begin{equation}
H_{obs}=-\dfrac{1}{(1+z)}{\left(\dfrac{dt}{dz}\right)}^{-1}\,,
\end{equation}
if approximated as in Eq, (\ref{ehsi}). The $\chi^2$ from the current $31$ OHD measurements reads
\begin{equation}
\chi^2_{OHD}=\sum_{i=1}^{31}\left[\dfrac{H_{th}({\rm x},z_i)-H_{obs}(z_i)}{\sigma_{H,i}}\right]^2\,.
\end{equation}
This procedure has the great advantage of directly considering $H$ without passing through any cosmic distance.

\subsection{The $\mathcal R$ parameter}

The CMB represents
cosmic recombination epoch remnant and contains abundant early universe information. Consequently, the acoustic peak positions \citep{Peebles70,Bond84} can be used to characterize a given cosmological model by means of the shift
parameter, defined as \citep{Bond97}
\begin{equation}
\label{shift}
\mathcal{R}=\frac{\sqrt{\Omega_M}}{\sqrt{|\Omega_k|}}{\rm
sinn}\left(\sqrt{|\Omega_k|}\int_{0}^{z_{\rm
ls}}\frac{dz}{E({\bf x},z)}\right)=1.70\pm 0.03\,.
\end{equation}
The last scattering redshift, namely $z_{\rm ls}$, is fixed to
\begin{equation}
z_{\rm ls}=1048[1+0.00124 (\Omega_{b}h^2)^{-0.738}][1+g_1
(\Omega_{M}h^2)^{g_2}],
\end{equation}
where $g_1=0.078(\Omega_b h^2)^{-0.238} [1+39.5 (\Omega_b
h^2)^{0.763}]^{-1}$ and $g_2=0.56 [1+21.1 (\Omega_b
h^2)^{1.81}]^{-1}$ \citep{Hu96} and the $\chi^2$ reads
\begin{equation}
\chi^{2}_{\rm CMB}=\frac{(\mathcal{R}-1.70)^2}{0.03^2}.
\end{equation}
In analogy with BAO measures, the shift parameter is not fully model-independent.

\subsection{Confidence levels and uncertainties }

As one performs fits combining GRBs with other observable quantities, meaningful information on the best-fit parameters are achieved by  computing their confidence limits or contour plots, which define the allowed parameter phase-space. These are essentially regions constructed around a set of best fit parameters got from computation. One does not mind about the number of dimensional parameter space, namely $m$, corresponding \emph{de facto} to the  number of parameters, since to make those regions compact one holds constant $\chi^2$ boundaries, fixing the chi squared values to specific numbers. So one takes $m$ to be the number of parameters, $n$ the number of data and $p$ be the confidence limit that one desires to reach. Assuming to shift by solving
$Q[n-m,min(\chi^2)+\Delta \chi^2]=p$, and to find the parameter region where $\chi^2\le min(\chi^2)+\Delta \chi^2$, immediately one gets the requested confidence region. Once the regions have been computed, it is needful to get uncertainties. To do so, expanding the log likelihood in Taylor series
$\ln {\cal L}=\ln{\cal L}(\theta_0)+\frac{1}{2}\sum_{ij}(\theta_i-\theta_{i,0})\left.\frac{\partial^2\ln{\cal L}}{\partial \theta_i \partial\theta_j}\right|_{\theta_0}(\theta_j-\theta_{j0})+...$,
we define the Hessian matrix by
\begin{equation}
{\cal H}_{ij}=-\frac{\partial^2\ln{\cal L}}{\partial \theta_i \partial\theta_j}.
\end{equation}
Since its non diagonal terms indicate correlated parameters, one can assume the errors on a given $i$ parameter to be  $1/\sqrt{{\cal H}_{ii}}$. This naive representation of errors is a coarse-grained approach, dubbed  conditional error, not frequently adopted in the literature. On the other hand, one can compute the Fisher  information matrix, as a forecast expression for error bars
\begin{equation}\label{fisher}
F_{ij}=\left\langle {\cal H}\right\rangle=-\left\langle \frac{\partial ^2 \ln {\cal L}}{\partial \theta_i \partial \theta_j}\right\rangle,
\end{equation}
with the ensamble average over observational data. In analogy to conditional errors, we write
$\sigma^2_{ij}\ge (F^{-1})_{ij}$, while
the marginalized errors become $
\sigma_{\theta_i}\ge(F^{-1})_{ii}^{1/2}$.

Above we underlined that the Fisher matrix is somehow related to error bars. In this respect, we mean the Fisher Information matrix enables to estimate the parameters errors before the experiment is performed. Hence, it permits to explore different experimental set ups that could optimize the experiment itself. For these reasons, the Fisher matrix is largely adopted in the literature.

\subsection{Binning procedure}

In several cases, it is useful to get constraints directly on the universe equation of state.
Thus, fitting it for the late-time universe constituents is extremely important to understand the dark energy evolution. In particular, pointing out a possible variation of the equation of state of dark energy is essential to \emph{disentangle} the standard model predictions from possible theoretical extensions and, in this respect, GRBs can be seen as intermediate redshift probes to disclose such an evolution.

To do that, an intriguing strategy consists in binning the dark energy equation of state, say $w$, in short intervals of  $z$ and then fit $w$ in each bin, assuming it is constant in each bin.
Indicating with a generic  function $f(z)$ the dark energy evolution, we have
\begin{equation}
  \label{eq:fzbinned}
  f(z_{n-1}<z \le z_n)=
  (1+z)^{3(1+w_n)}\prod_{i=0}^{n-1}(1+z_i)^{3(w_i-w_{i+1})},
\end{equation}
where $w_i$ is the barotropic factor within the $i^{\mathrm{th}}$ redshift
bin. The bin is built up by an upper boundary at $z_i$, whereas the zeroth bin is
defined as $z_0=0$.

There uncorrelated sub-equations of state in every bin can be experimentally refined adding data points and, in particular, GRBs, being calibrated as we will discuss later. Several indications have shown good agreement with the standard paradigm, up to $z\simeq 9$, albeit relevant deviations have been found, indicating the situation is not still clear.

\section{Standardizing GRBs}

Being successful in standardizing GRB data is of utmost importance to characterize new data catalogs up to high redshifts. In particular, getting redshifts, or more generally spectroscopic observations, is essential for GRB-related science, as we summarized below.
\begin{itemize}
\item[{\bf (1)}] Computing the luminosity function
for GRBs, constructing it from the prompt emission as well as
afterglows. This treatment is analogous to what we do for SNe Ia.
\item[{\bf (2)}] Computing the redshift distribution of
GRBs. This enables one to use GRBs as tracers for the cosmic star-formation history. Consequently, spotting very high redshift GRBs will shed light on their distribution at intermediate epochs of the universe evolution.
\item[{\bf (3)}] Studying the host galaxies, in particular those faint, high-redshift galaxies
that are unlikely to be found and studied with other methods. Characterising the dust extinction
curves of high-z galaxies.
\item[{\bf (4)}] Studying GRB-selected absorption line systems and probing cosmic chemical evolution
with GRBs.
\item[{\bf (5)}] Studying if and how much GRBs can be used for determining the cosmological parameters of dark energy models and/or to rule out a few models. Analogously the use of GRBs can be tested in view of determining cosmographic parameters, i.e. getting model independent bounds over the cosmic evolution.
\end{itemize}

\subsection{GRB correlations and related issues}\label{GRBcosmo}

Since the first discovery of GRBs independent groups have found
different correlations, that represent a key to use GRBs for cosmological purposes. The basic idea is to intertwine different quantities of such objects among them.
The observable quantities of interest are in relation with the cosmological model that lies on the background. This fact permits GRBs to be  distance indicators at a first glance, but limits their use because requires to postulate the underlying cosmological model, providing a circularity in the process itself which is know as the {\it circularity problem}.

The widest majority of GRB correlations prompt the same requirement: the GRB standardization in terms of cosmological tools.
Attempts for new correlations have been severely investigated, relating different observable quantities with each other. The way in which this is realized provides the  theoretical interpretation behind the relation itself. In other words, evidence for a given correlation leads to interpret particular physical processes. Thus, achieving the goal of standardizing GRBs brings the certainty of getting feasible bounds on cosmological parameters.  Intriguingly,  a narrow set of correlations enables one to  estimate GRB redshifts also. Even though this is still under speculation, in general a wide number of correlations could provide information about GRB progenitors.

More precisely, standardizing GRBs for cosmological purposes aims at reaching further hints toward  progenitors of \emph{different groups of GRBs}. Multi-wavelength instruments of recently-adopted satellites have significantly increased the number of GRBs that could be observed to check the validity of a given relation.
So, it is even possible a few correlations may be derived from experimental evidence, instead of theoretically. Unfortunately, this could open further issues related to data processing whose outputs can be biased in the overall computations.

Going ahead, it is certainly possible to constantly observe new hints undertaking
novel correlations to let free theoretical speculations that deeply search into new physics beyond the standard comprehension of GRBs.

\subsection{Prompt emission GRB correlations}

The correlations that make use of prompt emission quantities are listed below with the corresponding properties of each of them.  For details on the involved quantities, see Sec.~\ref{sec:2.5}.

\subsubsection{$L_{\rm iso}$--$\tau_{\rm lag}$ correlation.}

This correlation  holds for LGRBs and indicates that the more
luminous bursts possess shorter time lags, i.e., $L_{\rm iso}\propto \tau_{\rm lag}^{-1.25}$ \cite{2001AIPC..587..176N}.
It has been used as a GRB redshift indicator and to constrain cosmological parameters.
However, the existence of this correlation is challenged by recent studies. For details, see Ref.~\cite{2018AdAst2018E...1D}.

\subsubsection{
$L_{\rm iso}$--$V$ correlation.}

This correlation  holds for LGRBs and indicates that the more luminous bursts have the more variable light curves \cite{2000astro.ph..4176F}.
However, the intrinsic scatter is very large and the index is still not completely settled, also in view of the fact that the variability $V$ is different for various instruments. For details, see Ref.~\cite{2018AdAst2018E...1D}.

\subsubsection{
Amati or $E_{\rm p}$--$E_{\rm iso}$ correlation}

This correlation is of the form
$E_{\rm p}\propto E_{\rm
iso}^{0.52}$ and shows that $E_{\rm iso}$ is correlated with the rest-frame spectral peak energy, namely $
E_{\rm p} = E_{\rm p}^{\rm obs}(1+z)$ \cite{Amati2002}. Observations by \emph{Swift} and \emph{Fermi} detectors confirmed this correlation for LGRBs. An analogous $E_{\rm p}$--$E_{\rm iso}$ correlation with a slope similar to that of LGRBs but a larger value of the normalization holds also for SGRBs, though with much smaller data set \cite{2015MNRAS.448..403C}.
Moreover, the $E_{\rm p} - E_{\rm iso}$ correlation also holds within individual GRBs
using time-resolved spectra, and the slopes are consistent with the
correlation from time-integrated spectra. For details, see Ref.~\cite{2018AdAst2018E...1D}.
\subsubsection{Yonetoku or $L_{\rm p}$--$E_{\rm p}$ correlation}

This correlation reads $L_{\rm p}\propto E_{\rm p}^2$ \cite{Yonetoku2004} and holds for both LGRBs and SGRBs \cite{2015MNRAS.448..403C}. Similarly to the $E_{\rm p}$--$E_{\rm iso}$ correlation, the $L_{\rm p}$--$E_{\rm p}$ correlation holds also within individual GRBs using time-resolved spectra. For details, see Ref.~\cite{2018AdAst2018E...1D}.

\subsubsection{Ghirlanda or $E_{\rm p}$--$E_\gamma$ correlation}

This represents a tight, less scattered, correlation between $E_{\rm p}$ and $E_\gamma$, valid for LGRBs \citep{Ghirlanda2004}. One of the major drawbacks is the lack of achromatic breaks in the \emph{Swift} afterglow light curves of most of GRBs. This fact limits the increase in the correlation sample. For details, see Ref.~\cite{2018AdAst2018E...1D}.

\subsection{Prompt and afterglow emission correlations}

The following correlations involve prompt and afterglow emission observables. Below we report the most common correlations. For details on the involved quantities, see Sec.~\ref{sec:2.5}.

\subsubsection{Liang-Zhang or $E_{\rm p}$--$E_{\rm iso}$--$t_{\rm b}$ correlation}

It is a correlation valid for LGRBs among $E_{\rm iso}$, $E_{\rm p}$ and the rest-frame break time in the optical band $t_{\rm b}$, i.e., $E_{\rm p}\propto E_{\rm iso}^{0.52}t_{\rm b}^{0.64}$ \cite{2005ApJ...633..611L}.
If we take the optical break time as the jet break time, this correlation is similar to the Ghirlanda one. However, the inclusion of additional GRBs made this correlation more scattered. For details, see Ref.~\cite{2018AdAst2018E...1D}.

\subsubsection{Dainotti or $L_{\rm X}$--$t_{\rm X}$ correlation}

This correlation links the X-ray luminosity $L_{\rm X}$ and rest-frame time $t_{\rm X}$, i.e., $L_{\rm X}\propto t_{\rm X}^{-1}$, at the time when the X-ray afterglow light curve establishes a power-law decay after the plateau phase \cite{Dainotti2008}. This correlation holds for LGRBs and SGRBEEs. By adding a third parameter, $E_{\rm iso}$, it has been found the new correlation of the form
$L_{\rm X}\propto t_{\rm X}^{-0.87}E_{\rm iso}^{0.88}$ \cite{Dainotti2011}.
However, both relations are quite scattered and seem to be a selection effect due to the flux detection limit of {\em Swift}-XRT instrument. For details, see Ref.~\cite{2018AdAst2018E...1D}.

\subsubsection{
$E_{\rm iso}^{\rm X}$--$E_{\rm iso}$--$E_{\rm p}$ correlation}

This is a universal correlation for both LGRBs and SGRBs which links $E_{\rm iso}$ and $E_{\rm p}$ to the isotropic energy of the X-ray afterglow $E_{\rm iso}^{\rm X}$ computed in the rest-frame energy band $0.3$--$30$~keV, i.e., $E_{\rm iso}^{\rm X}\propto E_{\rm iso}^{1.00} E_{\rm p}^{0.60}$ \cite{Bernardini2012}. However, due to the fact that this correlation depends upon two cosmology-dependent quantities, it is unsuitable to constrain cosmological parameters. For details, see Ref.~\cite{Margutti2013}.
\subsubsection{
Combo correlation}

The correlation represents the combination of the Amati correlation and the $E_{\rm iso}^{\rm X}$--$E_{\rm iso}$--$E_{\rm p}$ correlation  and, like the Amati correlation, it holds for LGRBs only \cite{Izzo2015}.
It relates the prompt emission $E_{\rm p}$ and the X-ray afterglow  luminosity $L_0$ and rest-frame duration $\tau$ of the plateau phase, and the late power-law decay index $\alpha$, i.e., $L_0\propto E_{\rm p}^{0.90}\tau^{-1}|1+\alpha|$.
The main drawback is that it is much more complicated than other correlations, as it depends upon four parameters.
For details, see Ref.~\cite{2021ApJ...908..181M}.

\subsubsection{
$L$--$T$--$E$ correlation}

This correlation connects the rest-frame end time $t_{\rm X}$ and luminosity $L_{\rm X}$ of the X-ray afterglow plateau phase with $E_{\rm iso}$, {\it i.e.}, $L_{\rm X}\propto t_{\rm X}^{-1.01} E_{\rm iso}^{0.84}$ \cite{2019ApJS..245....1T}. This correlation is very similar to the Combo correlation but unlike it, the $L$--$T$--$E$ one holds for a few SGRBs and requires to  be corrected for the redshift evolution effects. For more details, see Ref.~\cite{2019ApJS..245....1T}.

\section{Further issues related to constructing GRB correlations}\label{sec:7}

Despite having so many correlations reported in the literature, all of them suffer of several issues related to constructing the correlations themselves. Besides the circularity problem, we face several issues to be addressed. In the following we list the most common ones.
\subsection{Evolution effects}

GRBs are observed from a large redshift range and in principle correlation parameters may
evolve with the redshift.
In some cases, to estimate the cosmological parameters a correction of the kind of $(1+z)^{-d}$ to the energy or luminosity parameters is needed, introducing a further parameter to fit.
However, though tested for subsamples of GRBs and several correlations, this issue is still ongoing \cite{2008MNRAS.391..411B}.
\subsection{
Instrumental selection effects}

A yet open issue are the instrumental selection effects that may affect the observed GRB energy or luminosity correlations.
There are at least two kinds of issues due to a) the trigger threshold, {\it i.e.}, the minimum photon peak flux that a burst must have in order to be detected by
a given instrument, and b) the spectral analysis threshold, {\it i.e.}, the minimum fluence to perform a reliable spectral analysis and determine the SED parameters.

Several analyses have been performed in the literature, using different samples and correlations of GRBs and searching for outliers and inconsistent GRBs.
Several conclusions were drawn: a) some correlations may exist, though due to selection effects; b) other correlations may exist when accounting for the intrinsic scatter; c) some correlations may have statistical significance, though affected by the thresholds of GRB detectors, etc.
See Ref.~\cite{2009AIPC.1133..350N}, for more details.
Interestingly, using the time-resolved spectra, similar correlations were found in individual bursts, strongly supporting the fact that the correlations may be physical \cite{2011MNRAS.418L.109G}.

\subsection{Systematic errors}

Sources of systematic errors for GRBs are the sensitivity of the detectors, the differences in the estimated spectral parameters depending on detectors and/or fitting models, the lack of unknown parameters, etc.
All these might dominate over the intrinsic dispersions of GRBs.

In general, due to the vast number of systematic errors, a technique to consider them in GRB fitting procedures consists in deriving those errors requiring the chi-square to be comparable to the fitted degrees of freedom  $\nu$, namely $\chi^2\simeq \nu$, and then summing them in quadrature with the statistical errors \cite{Schaefer2007}.

However, studies on different GRB correlations suggest that the systematic uncertainties of correlation parameters are not sensitive to the assumptions about cosmological parameters \cite{2008PhRvD..78l3532W}.

\subsection{Issues and interpretation of prompt emission GRB correlations}

In the following, attention is given to the above described GRB prompt emission correlations, focusing on possible issues and their physical interpretation.
Particular emphasis is given to the functional form of $E_{\rm p}$--$E_{\rm iso}$ (or Amati), Ghirlanda and Yonetoku correlations, the most popular and most quoted correlations for prompt emission observables.

\subsubsection{$L_{\rm iso}$--$\tau_{\rm lag}$ correlation.}

As established in Ref.~\cite{2010MNRAS.406.2149M}, this correlation holds also for X-ray
flares (in the rest frame
energy band $0.3$--$10$ keV) and proposed that their underlying mechanism is similar.
However, this correlation is affected by evolution effects, as discussed in Ref.~\cite{2018AdAst2018E...1D}.

One of the latest proposed explanations of the $L_{\rm iso}$--$\tau_{\rm lag}$ correlation involves only kinematic
effects \cite{2012ApJ...758...32S}, as the observed time-lag is $\tau_{\rm lag}\propto\mathcal{D}^{-1}$ and the luminosity is $L_{\rm iso}\propto\mathcal{D}$, where $\mathcal D$ is the Doppler boost defined in Sec.~\ref{sec:radiation}.

As discussed in Ref.~\cite{2019arXiv190306989S}, this correlation was constructed from a small sample of heterogeneously collected GRBs and is severely affected by sample incompleteness.

\subsubsection{
$L_{\rm iso}$--$V$ correlation.}

This correlation has a non-negligible scatter, thus, it is the least reliable one among all GRB prompt emission correlations \cite{2018AdAst2018E...1D}. The physical origin of the $L_{\rm iso}$--$V$ correlation is still unclear.
Within the internal shock scenario  (see Sec.~\ref{sec:scalings2}), it seems to be related to the activity of the central engine through the values of $\Gamma$ and the jet-opening angles \cite{2002ApJ...569..682S} and, based on this interpretation, $L_{\rm iso}$ ($V$) is proportional to a high (low) power of $\Gamma$, hence high-luminosity pulses imply high variability prompt light curves \cite{Schaefer2007}.

As discussed above for the $L_{\rm iso}$--$\tau_{\rm lag}$ correlation, the luminosity-variability correlation as well is severely plagued by sample incompleteness \cite{2019arXiv190306989S}.

\subsubsection{Amati or $E_{\rm p}$--$E_{\rm iso}$ correlation}

Likely, the most used and investigated relation is represented by the so-called $E_{\rm p}$--$E_{\rm iso}$ or Amati correlation \citep{AmatiDellaValle2013} that can be here recast by
\begin{equation}
\label{Amatirel}
\log\left(\frac{E_{\rm p}}{{\rm keV}}\right)= a_0 + a_1 \left[\log\left(\frac{E_{\rm iso}}{{\rm erg}}\right)-52\right]\,.
\end{equation}
Here, we have two free constants, namely $a_0$ and $a_1$, that represent the calibration constants to determine once the relation is somehow calibrated.
A possible limitation of the $E_{\rm p}$--$E_{\rm iso}$ correlation is due to the extra source of variability $\sigma_{\rm a}$. This is thought as a direct consequence of hidden variables that contribute to the overall calibration, albeit we cannot directly observe them \citep{Dago2005}.

A possible explanation for the $E_{\rm p}$--$E_{\rm iso}$ correlation considers the thermal radiation emitted when the GRB jet drills through the core of the progenitor star  (see Sec.~\ref{ssec:grbsn}), responsible for the thermal peak in the spectrum, and the Compton scattering of this radiation by relativistic electrons outside the photosphere  (see Sec.~\ref{sec:radiation} and Ref.~\cite{2007ApJ...666.1012T}, for details).

There are claims that the Amati correlation is caused by some selection effect of observations, rather than being an intrinsic property of GRBs \cite{Butler_2007,2019arXiv190306989S}.
However, there is a general consensus on the fact that the correlation is real \cite{2008MNRAS.387..319G,2008MNRAS.391..639N,2009A&A...508..173A}, though detector sensitivity affects the correlations and a weak fluence dependence may be larger than the statistical uncertainty and contribute to the dispersion of the correlation \cite{2010PASJ...62.1495Y,2011MNRAS.411.1843S}.

\subsubsection{Ghirlanda or $E_{\rm p}$--$E_\gamma$ correlation}

From theoretical and observational arguments in favor of the jetted nature of GRBs \cite{2013RSPTA.37120275H}, the radiated GRB energy can be corrected by means of the collimation factor $f=1-\cos\theta$, leading to $E_\gamma=f\,E_{\rm iso}$. In particular, the jet opening angle $\theta$ is evaluated at the characteristic time $t_{\rm b}$ for specific assumptions on the circumburst medium, that can be assumed  homogeneous \cite{Sarietal1999}.
The functional form adopted here for the Ghirlanda relation reads
\begin{equation}
\label{Ghirlandarel}
\log\left(\frac{E_{\rm p}}{{\rm keV}}\right)= b_0 + b_1 \left[\log\left(\frac{E_\gamma}{{\rm erg}}\right)-50\right]\,,
\end{equation}
in which, as usual, $b_0$ and $b_1$ are the two free constants, fixed by means of calibration. It behooves us the extra scatter,  $\sigma_{\rm b}$, to better constrain the relation itself.

The Ghirlanda correlation shares with the $E_{\rm p}$--$E_{\rm iso}$ one a similar physical interpretation  (see Sec.~\ref{sec:radiation} and Ref.~\cite{2007ApJ...666.1012T}, for details). This correlation takes into account also the jet correction in the computation of the GRB energy output  (see Sec.~\ref{progandquest}).

 The Ghirlanda correlation is linked to the Amati one and, thus, criticisms/analyses against/in favor of being an intrinsic property of GRBs. In Ref.~\cite{2005ApJ...627..319B} it was shown that as many as $33\%$ of the BATSE bursts would not be consistent with the Ghirlanda relation, but this results depended upon the assumed distribution for the jet's correction factor $f$ \cite{2005ApJ...627....1F}. This fact limits the increase in the correlation sample. For details, see Ref.~\cite{2018AdAst2018E...1D}.
Likewise for the Amati correlation, the Ghirlanda one is statistically real but
strongly affected by the thresholds of GRB detectors \cite{2011MNRAS.411.1843S}.

\subsubsection{Yonetoku or $L_{\rm p}$--$E_{\rm p}$ correlation}

The Yonetoku or $L_{\rm p}$--$E_{\rm p,}$ \cite{Yonetoku2004} correlation functional form here adopted reads

\begin{equation}
\label{Yonetokurel}
\log\left(\frac{L_{\rm p}}{{\rm erg/s}}\right) - 52 = m_0 + m_1 \log\left(\frac{E_{\rm p}}{{\rm keV}}\right).
\end{equation}
Here, the free terms are  $m_0$ and $m_1$ and require to be calibrated. Again, $\sigma_{\rm m}$ is the extra scatter term.

Detailed hydrodynamical simulations suggest that Yonetoku correlation may be due to the emission of photons from the photosphere of a relativistic jet, where the outflow becomes optically thin, whereas most of it is still optically thick (see Sec.~\ref{sec:radiation}). Quasi-thermal radiation is thus expected and the expected spectral shapes are obtained \cite{2019NatCo..10.1504I}.

 Like previous correlations, also for the Yonetoku correlation there are ongoing discussions whether it is a by-product of some selection effect or not \cite{Butler_2007,2008MNRAS.387..319G,2008MNRAS.391..639N,2009A&A...508..173A,2019arXiv190306989S}.
For this correlation, however, a weak redshift dependence has been confirmed, which may contribute to the dispersion of the correlation \cite{2010PASJ...62.1495Y}.

\subsection{Issues and interpretation of prompt and afterglow emission GRB correlations}

In analogy with the prompt emission correlations, we here focus on possible issues and physical interpretation of prompt and afterglow emission correlations.
The emphasis on the functional form is given to the Combo correlation, one of the less scattered correlations for both prompt and X-ray observables without evolution effects.

\subsubsection{Liang-Zhang or $E_{\rm p}$--$E_{\rm iso}$--$t_{\rm b}$ correlation}

The $E_{\rm p}$--$E_{\rm iso}$--$t_{\rm b}$ correlation \cite{2005ApJ...633..611L} has been proposed by purely considering phenomenological considerations,
thus avoiding any theoretical assumption, unlike made for the Ghirlanda correlation. However, as above stated, the $E_{\rm p}$--$E_{\rm iso}$--$t_{\rm b}$ somehow shares  similar implications and drawbacks than the Ghirlanda one, as well as analogous physical interpretation  (see Secs.~\ref{progandquest} and \ref{sec:radiation} and  Ref.~\cite{2007ApJ...666.1012T}, for details).

As proposed in Ref.~\cite{2012ApJ...745..168L}, this correlation, like the Ghirlanda one, appears to be affected by selection effect on $E_{\rm p}$ (whereas for the Ghirlanda correlation $E_\gamma$ is affected as well) and suffers sample incompleteness.

\subsubsection{Dainotti or $L_{\rm X}$--$t_{\rm X}$ correlation}

The Dainotti correlation, akin to the $L_{\rm iso}$--$\tau_{\rm lag}$ correlation, may be retrieved from kinematic effects  (see Sec.~\ref{sec:radiation} and Ref.~\cite{2012ApJ...758...32S}, fro details), pointing out to a common origin between the two correlations.

As already discussed above, this correlation is quite scattered and seem to be a selection effect due to the flux detection limit of {\em Swift}-XRT instrument \cite{2018AdAst2018E...1D}. Moreover, it is also affected by evolution effects with the redshift \cite{2017NewAR..77...23D}.

\subsubsection{
$E_{\rm iso}^{\rm X}$--$E_{\rm iso}$--$E_{\rm p}$ correlation}

This correlation depends upon two cosmology-dependent quantities, although  it is unsuitable to constrain cosmological parameters.
Since it holds for both SGRBs and LGRBs, with different progenitor and surrounding medium properties  (see Secs.~\ref{ssec:grbsn} and \ref{ssec:grbkn}), its physical interpretation has not been yet established.
There is a speculation that it may be connected with the $\Gamma$ of the outflow, which might regulate the efficiency of conversion from $\gamma$-rays to X-rays \cite{Bernardini2012}.

The $E_{\rm iso}^{\rm X}$--$E_{\rm iso}$--$E_{\rm p}$ correlation utilizes the prompt emission observables $E_{\rm iso}$ and $E_{\rm p}$ on which the Amati correlation is based. For this reason it is straightforward to deduce that the biases and selection effects, at work for the Amati correlation, partially affect this hybrid correlation. Moreover, unlike pure afterglow correlations such as the Dainotti one, this correlation is also plagued by double truncation in the flux limit, both in the prompt and X-ray afterglow emissions, making difficult the correction for any selection effect and the use as redshift estimators and cosmological tool (see discussions in Ref.~\cite{2017NewAR..77...23D}).

\subsubsection{Combo correlation}

The Combo correlation is a hybrid correlation linking the prompt emission $E_{\rm p}$ and the observable quantities determined from the X-ray afterglow light curve, i.e., among all the rest-frame $0.3$--$10$~keV plateau luminosity $L_0$, its rest-frame duration $\tau$, and the late power-law decay index $\alpha$ \cite{Izzo2015}.
For each GRB, $L_0$, $\tau$, and $\alpha$ can be obtained by fitting the rest-frame $0.3$--$10$~keV flare-filtered afterglow luminosity light curves with the function $L(t)=(1+t/\tau)^\alpha$.\footnote{The flare-filtered luminosity light curves are iteratively fitted with the above function: at every iteration, data points with the largest positive residual are discarded, until a final fit with a p-value $>0.3$ is obtained.}
The general expression is much more complicated than previous ones and reads
\begin{equation}
\label{Comborel}
\log \left(\frac{L_0}{{\rm erg/s}}\right) = k_0 + k_1 \log \left(\frac{E_{\rm p}}{{\rm keV}}\right) - \log \left(\frac{\tau/{\rm s}}{|1+\alpha|}\right).
\end{equation}
Here, the constants $k_0$ and $k_1$ need to be determined by means of the calibration procedure. Again, the correlation is characterized by an extra scatter $\sigma_{\rm k}$.

The Combo correlation can be explained by the external shock scenario (see Sec.~\ref{progenitor}).
The correlation is the result of the synchrotron emission from the electrons accelerated in a relativistic shock  (see Sec.~\ref{sec:radiation}). The shock propagates through the external CBM and interacts with the
magnetic field of the the turbulent plasma. Hence, the relationship among $E_{\rm p}$, $L_0$ and $\tau$ and the corresponding comoving quantities scale with the initial Lorentz factor of the bulk motion $\Gamma_0$, whereas the intrinsic scatter is due to the uncertainties on the source spectral energy distribution \cite{2017MNRAS.468..570M}.

Keeping in mind the hybrid nature and the Combo correlation, which a combination of the Amati and the $E_{\rm iso}^{\rm X}$--$E_{\rm iso}$--$E_{\rm p}$ correlations, the same biases and selection effects at work in the $E_{\rm iso}^{\rm X}$--$E_{\rm iso}$--$E_{\rm p}$ affect the Combo correlation as well.

\subsubsection{
$L$--$T$--$E$ correlation}

As already stated, this correlation shares similarities with the Combo correlation but, unlike it, needs to be corrected for the redshift evolution effects \cite{2019ApJS..245....1T}.

The $L$--$T$--$E$ correlation, as well as the Combo correlation,  may be explained within the magnetar scenario, which justifies the plateau phase observed in the X-ray afterglow light curves as due to the continuous energy supply from a supra-massive NS  (see Sec.~\ref{sec:magnetized_flow} and Ref.~\cite{Bucciantini2008}, for details).

 Likewise the Combo correlation, which share similar features, the discussion on the same discussion on the biases and selection effects holds for the $L$--$T$--$E$ correlation as well. In addition, in Ref.~\cite{2019ApJS..245....1T} it is evidenced that GRBs of the sample at high redshifts usually have relatively larger $L_{\rm X}$ and $E_{\rm iso}$. This is very likely a selection effect due to the fact that very distant GRBs with small $L_{\rm X}$ and $E_{\rm iso}$ cannot be observed in view of the limited threshold of current detectors.

\section{Circularity or not circularity? }

Even though GRBs could be  thought as indicators toward the determination of the universe's expansion history, it is remarkable to stress that small redshift GRB data are still missing. Consequently if one requires to calibrate GRBs with small redshift data, there is the strict need of other data sets, whose data points lie around $z\simeq0$, that permit to perform the calibration procedure. The calibration procedure is essentially a consequence of the correlation functional forms that, by virtue of the above considerations, are commonly written as $y = a\, x + b$, with $a,b$ real constants that depend upon the relation itself. Bearing this in mind, we  focus below on different calibration strategies. In this respect, we  highlight whether it is absolutely necessary or not to calibrate our correlations in order to confront cosmological models with GRB data. Below we start with such considerations, critically discussed.

\subsection{Calibration versus non-calibration}

Constraining cosmological parameters
using GRBs is plagued by several conceptual and practical issues.

First, all GRBs correlations by definition are built assuming an \textit{a priori} background cosmology  (see Secs.~\ref{sec:2.5} and \ref{GRBcosmo} for details) and, consequently, introducing a \emph{circularity problem} \cite{Kodama2008}.
Second, almost the majority of GRB correlations holds for LGRBs  (see Sec.~\ref{GRBcosmo} for details), whose observational rate falls off rapidly at low-$z$ and in some cases such a nearby LGRBs seems to be intrinsically different from the \textit{cosmological} ones \cite{Galama1998,Soderberg2008Nature}.
Third, unlike SNe Ia which are calibrated with a selected sub-sample at very low redshift\footnote{Where their luminosity distance is essentially independent from the choice of the cosmological model} by anchoring them to primary distance indicators as Cepheids, the shortage of low-$z$ GRBs prevents to anchor them to primary distance indicators.

Focusing on the above \emph{circularity problem}, this is essentially an epistemological issue due to the lack of very low-$z$ GRBs and arising from the need of a background cosmology to compute the above-defined
$E_{\rm iso}$, $E_\gamma$ and
$L_{\rm iso}$ entering GRB correlations \cite{Kodama2008}.
For example, calibrating GRBs through the standard $\Lambda$CDM model, the estimate of cosmological parameters of any dark energy framework inevitably returns an overall agreement with the concordance model.
Debates toward its use in cosmology seem to indicate that this effect could be minor, albeit it plagues cosmological constraints got from GRBs.

Possible way outs known in the literature, involve calibration techniques based on the use of SNe Ia distance moduli, cosmographic series, cosmic chronometers, etc. All these procedures represent plausible solutions, but at the same time introduce possible issues that we are going to describe below.

The above calibration procedures have to compare with the an alternative method which completely by-passes the calibration procedure \cite{2021arXiv210512692K}.
This uncalibrated procedure consists in a simulataneous fit of correlation parameters together with the cosmological model parameters. This uncalibrated procedure consists in a simultaneous fit of correlation and  cosmological model parameters.
This procedure and the related issues are also discussed in the following.

\subsection{Fitting procedures with calibration}

\subsubsection{SN calibration}

A widely-used method to calibrate GRB correlations is through the use of SNe Ia, that span within $z\lesssim 2.3$.
In such a way, assuming this could work for any LGRBs, the GRB data points are mixed with  SNe in order to build up a whole, quite large,
Hubble diagram, where in the small redshift domain one has the majority of SNe, while at large $z$, GRBs are the most.
Here, the simplest error bars on  distance modulus are \citep{Liang08,Wei10}
\begin{equation}
\sigma_{\mu}=([(z_{i+1}-z)/(z_{i+1}-z_i)]^2\epsilon_{\mu,i}^2+[(z-z_{i})/(z_{i+1}-z_i)]^2\epsilon_{\mu,i+1}^2)^{1/2},
\end{equation}
where $\epsilon_{\mu,i}$ and $\epsilon_{\mu,i+1}$ and $\mu_{i}$ and  $\mu_{i+1}$ are the errors and distance moduli of the SNe Ia at $z_{i}$ and $z_{i+1}$, respectively.

For each SN catalog, we could find different interpolating functions to model the SN distribution. Thus, calibrating GRBs with SNe would
seriously depend on the choice of these expressions for each catalog. Hence, GRB calibrations may turn out to be extremely sensitive to SNe Ia and the approach should be carefully handled since
GRB luminosity correlations may be no longer fully independent from SN data points.

\subsubsection{Model dependent calibration}

The model dependent procedure fixes the background cosmology with a given cosmological model, where typically the dark energy evolution is assumed \emph{a priori}.
Since the background cosmology\footnote{The cosmological model under exam, or background cosmology, is intimately determined by knowing the functional form and evolution of $H(z)$.} enters the correlation functions, generically in the form $y=ax+b$, this means that one has to 1) assume a background cosmology, 2) fix the most suitable numerical bounds over the free coefficients of the background cosmology, and 3) calibrate the correlation.

As it appears evident, this strategy consists in determining an accredited cosmological model with particular choices of the free parameters, determined elsewhere.

This procedure is obviously strongly plagued by the circularity problem. It fixes the cosmological evolution with a given model and does not permit to constrain suitably another cosmological paradigm.
In fact, if one calibrates with a generic model, say $H^{(1)}(z)$, any other statistical expectations on a different model, say $H^{(2)}(z)$, would favor the model that better matches $H^{(1)}(z)$. In other words, calibrating with $H^{(1)}(z)$ implies that the best fits are statistically argued for $H^{(2)}(z)\simeq H^{(1)}(z)$.

Another dramatic fact is that one has to constrain the free parameters of the background scenario by means of additional fits, with different data sets. This implies that an overall analysis would be plagued by error propagation between different catalogs of data and limits severely the analysis itself. To avoid other fits, one can assume exact versions of the cosmological models that should be used as backgrounds. In such a way, the corresponding error propagation reduces, albeit one does not take a real tested cosmological scenario, but rather a simplified version of it.

\subsubsection{Model independent calibration}

Calibrating correlations via model independent treatments permits to use GRBs as distance indicators, although the calibration is made by means of other standard candles. The idea of model-independent calibrations, however, enables to get the luminosity distance $d_{\rm L}$ without \emph{a priori} postulating the background cosmology, healing \emph{de facto} the circularity problem.

A nice possibility consists of relating distances with model-independent quantities written in terms of a Taylor series expansion of the scale factor. Thus, we first notice
\begin{equation}
H(z)=\left\{\frac{d}{dz}\left[\frac{d_{\rm L}(z)}{1+z}\right]\right\}^{-1}\,,
\end{equation}
and then we consider the following expansions
\begin{align}
\label{dlandH}
d_L(z)&=\frac{z}{H_0}\sum_{n=0}^{N}\frac{\alpha_n}{n!}z^n\,,\\
\label{dlandH2}
H(z)&=\sum_{m=0}^{M}\frac{H_{m}}{m!}z^m\,,
\end{align}
where $\alpha_n$ are the coefficients of the luminosity distance expansion and $H_m$ are the coefficients of the Hubble rate expansion.
Thus, baptizing the cosmographic set, $q_0,j_0,s_0$, as the present values of the following quantities
\begin{align}\label{equazionissime}
q(t) = - \frac{1}{a} \; \frac{d^2 a}{d t^2}  \;\left[ \frac{1}{a} \;
\frac{d a}{d t}\right]^{-2},\quad
j(t) = + \frac{1}{a} \; \frac{d^3a}{d t^3}  \; \left[ \frac{1}{a}
\; \frac{d a}{d t}\right]^{-3},\quad
s(t) = + \frac{1}{a}, \frac{d^4a}{d t^4}\left[ \frac{1}{a}
\; \frac{d a}{d t}\right]^{-4}\,,
\end{align}
where the scale factor $a$ has been considered, with the requirement $a\equiv(1+z)^{-1}$. The above quantities are named {\it  deceleration}, {\it jerk}, and {\it snap} parameters, respectively, we formally have
\begin{equation}
d_{\rm L}^{(4)}\simeq\frac{z}{H_0}\left(\alpha_0+\alpha_1z+\alpha_2\frac{z^2}{2}+\alpha_3\frac{z^3}{6}\right)\,.
\end{equation}
The coefficients in Eqs.~\eqref{dlandH}--\eqref{dlandH2}, say $\alpha_i\equiv\alpha_i(q_0,j_0,s_0)$ and $H_i\equiv H_i(q_0,j_0,s_0)$, can be determined directly with data, without considering a cosmological model \emph{a priori}. This treatment is known with the name of  {\em cosmography} or {\em cosmokinetics}, {\it i.e.}, the part of cosmology  that reconstructs the universe's kinematics model-independently.
So, at $z=0$ we have
\begin{align}
\alpha_1&=\frac{1}{2}(1-q_0)\,,\\
\alpha_2&= -\frac{1}{6}(1-q_0-3q_0^2 +j_0)\,,\\
\alpha_3&=\dfrac{1}{24}(2 - 2 q_0 - 15 q_0^2 - 15 q_0^3 + 5 j_0 +10 q_0 j_0 +s_0)\,,
\end{align}
and
\begin{align}
H_1&=1 + q_0\,,\\
H_2&=\dfrac{1}{2} (j_0 - q_0^2)\,, \\
H_3&=\dfrac{1}{6}\left[-3 q_0^2 - 3 q_0^3 + j_0 (3 + 4 q_0) + s_0\right]\,.
\end{align}

Although powerful, the above formalism suffers from shortcomings due to the convergence at higher redshifts\footnote{For the sake of completeness, this problem is not related to GRB redshifts only, but it remains an open issue of cosmography.}, {\it i.e.}, the high GRB redshifts are very far from $z=0$.
In other words, the standard cosmographic approach fails to be predictive if one employs data at higher redshift domains, which is exactly the case of GRBs.

Healing the convergence problem leads to a high-redshift cosmography. In this respect, several strategies have been suggested. For instance, one could 1) extend the  limited convergence radii of Taylor series by changing variables of expansion, using the so-called \emph{auxiliary variables}, or 2) changing the mathematical technique in which the expansions are performed, {\it i.e.} involving expansions different from Taylor ones, etc.

In the case of auxiliary variables, one employs a tricky method in which the expansion variable is reformulated as a function of the redshift itself, but with particular convergence properties. In other words, we cosmic quantities are rewritten in a more complicated function of the redshift $z$, namely $y$.
Changing the redshift variable from $z$ to $y$ modifies accordingly the convergence radius. Formally speaking we write   $y\equiv\mathcal{F}(z)$ \cite{2019IJMPD..2830016C}, where we assume $\mathcal{F}(z)$ a generic function of the redshift.
The function $\mathcal{F}(z)$ is properly chosen from physical prime principles. All $\mathcal{F}(z)$ prototypes, however, might fulfill a few mathematical conditions:
\begin{itemize}
    \item[{\bf 1.}]  $\mathcal F(z)\big|{\,}_{z=0}=0$\,,
    \item[{\bf 2.}]  $\mathcal F(z)\big|{\,}_{z=0}<\infty$\,.
\end{itemize}
The first guarantees that at $z=0$, our time, even $y$ is zero. The second that the auxiliary variable does not diverge, otherwise the convergence problem would still persist.
In addition, a further  requirement is helpful in constructing $\mathcal{F}(z)$:
\begin{itemize}
    \item[{\bf 3.}] $\mathcal F(z)\big|{\,}_{z=-1}<\infty$\,.
\end{itemize}
The former condition enables $y$ to converge at future time, as well as $z$. Using these hints toward the formulation of $\mathcal{F}(z)$, we can suggest a couple of well-consolidate examples of $\mathcal{F}(z)$:
\begin{align}
y_1&=\frac{z}{1+z}\,,\\
y_2&=\arctan{z}\,.
\end{align}

The second possibility is offered by rational approximations, where the expansion is thought to be a rational function, instead of a polynomial. This guarantees  to optimize the Taylor series with rational approximants that better approach large $z$ than the Taylor series, guaranteeing mathematical stability of the new series if data points exceed $z=0$. Among all the possible choices, here the attention is given to the Pad\'e polynomials, firstly introduced in Ref.~\citep{2014PhRvD..89j3506G}.
This technique of approximations turns out to be a bookkeeping device to keep the calculations manageable for the cosmography convergence issue.
Thus, provided we have Taylor expansions of $f(z)$ under the form $
f(z)=\sum_{i=0}^\infty c_i z^i$, with  $c_i=f^{(i)}(0)/i!$, it is possible to obtain the $(n,m)$ Pad\'e approximant by
\begin{equation}
P_{n,m}(z)=\left(\sum_{i=0}^{n}a_i z^i\right)\left(1+\sum_{j=1}^{m}b_j z^j\right)^{-1}\,,
\label{eq:def Pade}
\end{equation}
and requires that $b_0=1$. Furthermore, it is important that $
f(z)-P_{n,m}(z)=\mathcal{O}(z^{n+m+1})$ and the coefficients $b_i$ come from  solving the homogeneous system of linear equations $
\sum_{j=1}^m b_j\ c_{n+k+j}=-b_0\ c_{n+k}$,
valid for $k=1,\hdots,m$. Once $b_i$ are known, $a_i$ can be obtained using the formula
$a_i=\sum_{k=0}^i b_{i-k}\ c_{k}$. Just for an example, we report the $(2,1)$ Pad\'e polynomial as
\begin{equation}
P_{2,1}(z)=\frac{z}{H_0}\left\{\frac{6 (q_0-1) + \left[q_0 (8 + 3 q_0) -5 - 2 j_0\right] z}{2 q_0 (3 + z + 3 q_0 z)-2 (3 + z + j_0 z)}\right\}\,.
\end{equation}

\subsection{The use of B\'ezier polynomials}

The left term of Eq. \eqref{ehsi} can be approximated by means of particular choices, such as using model-independent B\'ezier parametric curves. They are constructed to be stable at the lower degrees of control points. They can be rotated and translated by performing the operations on the points and assuming a degree $n$.
They formally are defined as
\begin{equation}
\label{bezier}
H_n(z)=\sum_{d=0}^{n} \beta_d h_n^d(z)\quad,\quad
h_n^d(z)\equiv \frac{n!(z/z_{\rm m})^d}{d!(n-d)!} \left(1-\frac{z}{z_{\rm m}}\right)^{n-d}\,,
\end{equation}
where we notice the linear combination of Bernstein basis polynomials $h_n^d(z)$. Assuming the coefficients $\beta_d$ of to be positive in the range of $0\leq z/z_{\rm m}\leq1$, where $z_{\rm m}$ is the maximum $z$ of OHD, we soon can classify those polynomials by means of the exponent $n$.

In particular, besides the  constant case, $n=0$, both linear growth, that happens for $n=1$ and oscillatory regimes, say $n>2$, work well. This implies that a suitable choice is $n=2$. In this case we have
\begin{equation}
\label{bezier2}
H_2(z)=\beta_0\left(1-\frac{z}{z_{\rm m}}\right)^2 + 2 \beta_1 \left(\frac{z}{z_{\rm m}}\right)\left(1-\frac{z}{z_{\rm m}}\right) + \beta_2 \left(\frac{z}{z_{\rm m}}\right)^2\,.
\end{equation}
The comparison between $H_2(z)$ and the OHD data points give $\beta_0=67.76\pm3.68$, $\beta_1=103.3\pm11.1$, and $\beta_2=208.4\pm14.3$, all in units of km~s$^{-1}$~Mpc$^{-1}$.

After having approximated $H(z)$ with Eq.~\eqref{bezier2}, for spatially flat cosmology,  $\Omega_k=0$,  the calibrating luminosity distance becomes
\begin{equation}
\label{dlHz2}
d_{\rm cal}(z)\simeq (1+z)\int_0^z\dfrac{dz'}{H_2(z')}\,.
\end{equation}
Once the luminosity distance is written, it is possible to calibrate $E_{\rm iso}$, $E_\gamma$, $L_{\rm p}$, and $L_0$ for the correlations that we intend to test.
For $E_{\rm p} - E_{\rm iso}$, Ghirlanda, Yonetoku and Combo correlations we report in Table~\ref{tab:1} the corresponding numerical outcomes related to the calibration process.
\begin{table*}
\setlength{\tabcolsep}{1.5em}
\renewcommand{\arraystretch}{1.1}
\centering
\caption{For brevity we report in this table only a few calibrated correlations. In particular, in the columns we prompt four correlation, {\it i.e.}, Amati, Ghirlanda, Yonetoku and Combo, with the data set number points and the corresponding last update year. On the right, we display the calibrated best fit parameters. The statistical method behind these  calibrations is reported in Sect. \ref{10.2}.}
\begin{tabular}{lcc|lll}
\hline\hline
Correlation         &   $N$     &   Update  & \multicolumn{3}{c}{Parameters}\\
\hline
\textit{Amati}      &   $193$   &   $2015$  &   $a_0=2.06\pm0.03$
                                            &   $a_1=0.50\pm0.02$
                                            &   $\sigma_{\rm a}=0.20\pm0.01$\\
\textit{Ghirlanda}  &   $27$    &   $2007$  &   $b_0=2.09\pm0.04$
                                            &   $b_1=0.63\pm0.04$
                                            &   $\sigma_{\rm b}=0.10\pm0.02$\\
\textit{Yonetoku}   &   $101$   &   $2018$  &   $m_0=-3.43\pm0.21$
                                            &   $m_1=1.51\pm0.08$
                                            &   $\sigma_{\rm m}=0.35\pm0.03$\\
\textit{Combo}      &   $60$    &   $2015$  &   $q_0=50.04\pm0.27$
                                            &   $q_1=0.71\pm0.11$
                                            &   $\sigma_{\rm q}=0.35\pm0.04$\\
\hline\hline
\end{tabular}
\label{tab:1}
\end{table*}

Once calibrated, the corresponding distance moduli from Eq.~\eqref{defmu} are computed for each correlations.

\subsubsection{Simultaneous fits}

Another relevant strategy is based on the idea to constrain the cosmological parameters \emph{together with} the luminosity correlation  \cite{,Ghirlanda04b,Wang07}.
In particular, the real distance modulus can be computed as
\begin{equation}
\mu_{\rm fit}=\frac{\sum_i \mu_i/\sigma_{\mu_i}^2}{\sum_i
\sigma_{\mu_i}^{-2}},
\end{equation}
where the sum is over a given number of different correlations. In particular, $\mu_i$ is the best estimated distance modulus and the subscript $i$-th refers to the correlation, with $\sigma_{\mu_i}$  the error bars. The uncertainty of the distance modulus for each burst is $
\sigma_{\mu_{\rm fit}}=(\sum_i \sigma_{\mu_i}^{-2})^{-1/2}$.

A great advantage is that, as one computes bounds on cosmological parameters, the normalization functions and slopes of each correlations are marginalized. Consequently, we write down the $\chi^2$ as
\begin{equation}
\chi^{2}_{\rm GRB}=\sum_{i=1}^{N}
\frac{[\mu_{i}(z_{i},H_{0},\Omega_{M},\Omega_{DE})-\mu_{{\rm
fit},i}]^{2}}{\sigma_{\mu_{{\rm fit},i}}^{2}},
\end{equation}
where $\mu_{{\rm fit},i}$ and $\sigma_{\mu_{{\rm fit},i}}$ are the
fitted distance modulus and its error respectively.

\subsubsection{Narrow calibration}

Another intriguing technique consists in calibrating standard candles using GRBs in a narrow redshift range, hereafter $\delta z$. This short interval is placed near a fiducial redshift
\cite{Liang2006,Ghirlanda06b} with the great advantage that, in some cases, see e.g. Ref. \cite{Liang2006}, no low-redshift GRB sample is
needful.

\subsection{Fitting procedures without calibration}

Constraints on the cosmological parameters can be obtained with an alternative method which completely by-pass the calibration procedure.
It consists in taking all best data set of any GRB correlations, introducing the lowest intrinsic dispersion. The method is described below.

Any correlation between a generic energy/luminosity quantity $\mathcal{Y}$ and a GRB observable $\mathcal{X}$ has the form
\begin{equation}
\label{eq:correlation}
\log\mathcal{Y}^{\rm obs} = a\log\mathcal{X} + b\,.
\end{equation}
The energy/luminosity quantity in general contains the information on the cosmological parameters $\Omega_{\rm i}$ through $d_{\rm L}(z,\Omega_{\rm i})$, defined by the theoretical model describing the background cosmology
\begin{equation}
\label{eq:correlation2}
\mathcal{Y}^{\rm th} = 4\pi d^2_{\rm L}(z,\Omega_{\rm i}) \mathcal{F}_{\rm bolo}\,,
\end{equation}
where $\mathcal{F}_{\rm bolo}$ may be the rest-frame bolometric fluence $S_{\rm bolo}(1+z)^{-1}$ for Amati-like correlations, or the bolometric observed flux $F_{\rm bolo}$ for Yonetoku-like correlations. Please, notice that we only focus on these two relations just for giving an example. The same can be reformulated for other correlations, although for brevity we do not report other treatments here.

The best cosmological and correlation parameters are then obtained by maximizing the log-likelihood function \citep{Dago2005}
\begin{equation}
\label{eq:chi2}
    \ln\mathcal{L} = -\frac{1}{2}\sum^{N}_{i = 1} \left[\frac{\left(\log\mathcal{Y}^{\rm obs}_{\rm i} - \log\mathcal{Y}^{\rm th}_{\rm i}\right)^2}{\sigma^2_i} + \ln(2\pi \sigma^2_i)\right],
\end{equation}
where $\sigma^2_i = \sigma^2_{\log\mathcal{Y}^{\rm obs}_{\rm i}} + a^2 \sigma^2_{\log\mathcal{X}_{\rm i}} + \sigma^2_{\rm ext}$. Here, $\sigma_{\log\mathcal{Y}^{\rm obs}_{\rm i}}$ is the error in the measured value of $\log\mathcal{Y}^{\rm obs}_{\rm i}$, $\sigma_{\log\mathcal{X}_{\rm i}}$ is the error in $\log\mathcal{X}_{\rm i}$, and $\sigma_{\rm ext}$ is the intrinsic dispersion of the correlation.

The above treatment avoids to calibrate GRB data.
This procedure has been applied to an uncalibrated $E_{\rm p}$--$E_{\rm iso}$ correlation in Ref.~\cite{2020MNRAS.499..391K} and to a uncalibrated Combo correlation in Ref.~\cite{2021arXiv210512692K}. In both the cases, the correlations have been built up from samples of GRBs that have lower intrinsic dispersion.
As byproducts, the resulting GRB correlations are close for different cosmological models, which can be interpreted with the fact that this procedure is model-independent.
However, the application of this method seems to indicate that current GRB data are not able to put stringent constraints on cosmological parameters, though consistent with those resulting from better-established cosmological probes.

However, as hinted by the above results, this method may introduce a possible bias, namely that GRB correlation may adjust itself to the cosmology that maximize Eq.~\eqref{eq:chi2}, rather than allowing the derivation of $\Omega_{\rm i}$ from a cosmology independent calibrated correlation.
Moreover, it may be possible that more exotic cosmological model would lead to best-fit GRB correlations significantly different from simpler model, thus failing in providing a model-independent procedure.
Therefore, this method is yet the subject of ongoing studies.


\section{Recent developments of cosmology with gamma-ray bursts}\label{sec:4}

\subsection{Numerical results using correlations}

Observational data indicate that the cosmological expansion is currently accelerating. They also indicate that in the recent past the expansion was decelerated. The standard spatially-flat $\Lambda$CDM model \citep{Peebles1984,DiValentinoetal2021b,PerivolaropoulosSkara2021} is the simplest model consistent with these observations \citep{Farooqetal2017,Scolnicetal2018, PlanckCollaboration2020, eBOSSCollaboration2020}. Here, a cosmological constant $\Lambda$ dominates the current energy budget and fuels the currently-accelerating cosmological expansion. In this model, above a redshift $z\approx0.75$, non-relativistic cold dark matter and baryons dominate over $\Lambda$ and powered the then-decelerating cosmological expansion. While the observations are consistent with dark energy being time- and space-independent, they do not rule out slowly-evolving and weakly spatially-inhomogeneous dynamical dark energy, nor spatial flatness.

Significant constraints on cosmological parameters come from the CMB anisotropy data --- that primarily probe the $z \sim 1100$ part of redshift space --- as well as from BAO observations --- the highest of which reach to $z \sim 2.3$ --- and other lower-redshift SNeIa and OHD measurements. Observational data in the intermediate redshift range, between $z \sim 2.3$ and $\sim 1100$, are not as constraining as the lower and higher redshift data, but hold significant promise.
Intermediate redshift observations include those of HII starburst galaxies that reach to $z \sim 2.4$ \citep{ManiaRatra2012, Chavezetal2014,GonzalezMoran2019,GonzalezMoranetal2021,Caoetal2020,Caoetal2021a,Johnsonetal2021}, quasar angular sizes that reach to $z \sim 2.7$ \citep{Caoetal2017, Ryanetal2019, Caoetal2020, Caoetal2021b, Zhengetal2021, Lianetal2021},  quasar X-ray and UV fluxes that reach to $z \sim 7.5$ \citep{RisalitiLusso2015, RisalitiLusso2019, KhadkaRatra2020a, KhadkaRatra2020b, KhadkaRatra2021, Yangetal2020, Lussoetal2020, Lietal2021, Lianetal2021}, as well as GRBs, that have now been detected to $z=9.4$ \citep{Cucchiara2011}. Observed correlations between GRB photometric and spectroscopic properties that can be related to an intrinsic burst physical property would allow GRBs to be used as valuable standard candles that reach  high $z$ and probe a largely unexplored region of cosmological redshift space \citep[see e.g.][and references therein]{Schaefer2007, Amati2008, CapozzielloIzzo2008, Dainotti2008, Izzo2009, AmatiDellaValle2013, Wei2014, Izzo2015, 2019ApJS..245....1T}, similar to how SNeIa are used as standard candles \citep{Phillips1993} at $z < 2.3$. However, as stressed many times in this review, this is still a challenge for GRBs.

After it was established that GRBs were at cosmological distances, many attempts have been made to use burst correlations to constrain cosmological parameters. The first GRB Hubble diagram of a small sample of $9$ bursts, obtained in Ref.~\cite{2003ApJ...583L..67S} from the $L_{\rm iso}$--$V$ correlation \cite{2000astro.ph..4176F}, led to a current non-relativistic matter energy density parameter limit of $\Omega_{m0} < 0.35$ at the 1$\sigma$ confidence level (for the flat $\Lambda$CDM model). Soon after, using the Ghirlanda correlation, in Ref.~\cite{2004ApJ...612L.101D}, with a sample of $12$ bursts, it has been found $\Omega_{m0}=0.35\pm0.15$ for the flat $\Lambda$CDM model, and in Ref.~\cite{2004ApJ...613L..13G}, with $14$ GRBs as well as SNeIa, it has been inferred $\Omega_{m0}=0.37\pm0.10$ and a cosmological constant energy density parameter $\Omega_\Lambda=0.87\pm0.23$, in the non-flat $\Lambda$CDM model, and $\Omega_{m0}=0.29\pm0.04$ in the flat model. Similar constraints were obtained in Ref.~\cite{2005ApJ...633..611L}, using the $E_{\rm p}$--$E_{\rm iso}$--$t_{\rm b}$ correlation: $0.13 < \Omega_{m0} < 0.49$ and $0.50 < \Omega_\Lambda < 0.85$ at 1$\sigma$ confidence level in the flat $\Lambda$CDM model.

More recently, many contrasting results have been reported in the literature.
In Ref.~\cite{SamushiaRatra2010} cosmological parameter constraints, within the $\Lambda$CDM model and dynamical dark energy models from two different GRB data sets, were found to be different from the two data sets and also relatively broad. Similarly, in Ref.~\cite{LiuWei2015} it has been shown that at that time GRB data could not significantly constrain cosmological parameters. In addition, in Refs.~\cite{Linetal2016,Huangetal2020} it has been shown that most GRB correlations have large scatter and/or their parameters differ somewhat significantly between low- and high-$z$ GRB data sets. From the calibration of the Ghirlanda correlation, by using a SNeIa distance-redshift relation --through the  $(3,2)$ Pad\'e approximant--, in Ref.~\cite{Linetal2016} it has been obtained $\Omega_{m0}= 0.302\pm0.142$ within the flat $\Lambda$CDM model.\footnote{Similar conclusions have been reached in Ref.~\cite{Tangetal2021}, who confirmed that only the Ghirlanda correlation has no redshift dependence, and determined $\Omega_{m0} = 0.307^{+0.065}_{-0.073}$ in the flat $\Lambda$CDM model from SNeIa calibrated GRB data.}
Based on a cosmographic approach, an updated $E_{\rm p}$--$E_{\rm iso}$ correlation with $162$ GRBs has been used to get cosmological constraints. In Ref.~\cite{Demianskietal2017a} GRBs were calibrated with SNeIa, resulting in $\Omega_{m0}=0.25_{-0.12}^{+0.29}$ within the flat $\Lambda$CDM model, whereas in Ref.~\cite{Demianskietal2017b} a cosmographic expansion, up to the fifth order, involving SNeIa is used to calibrate the $E_{\rm p}$--$E_{\rm iso}$ correlation for GRBs, which are then used in conjunction with OHD and BAO measurements to constrain cosmographic parameters, resulting in a 1$\sigma$ deviation from the $\Lambda$CDM cosmological model.

Other recent works (involving GRB data only or in conjuction with other probes) also report inconsistencies with the $\Lambda$CDM model. In Ref.~\cite{Demianskietal2019} the $E_{\rm p}$--$E_{\rm iso}$ correlation has been used, including also modeling the potential evolution of GRB observables, to conclude that calibrated GRB, SNeIa, and OHD data favor a dynamical dark energy model described by a scalar field with an exponential potential energy density.
In Ref.~\cite{orlando3}, Amati, Ghirlanda, Yonetoku and Combo correlations have been calibrated in a model-independent way via OHD and jointly analyzed with SNeIa and BAO by using cosmographic methods, such as Taylor expansions, auxiliary variables and Padé approximations, to conclude that GRB do not favor the flat $\Lambda$CDM model but instead favor a mildly evolving dark energy density model.
Similarly, in Ref.~\cite{LM2020} the $E_{\rm p}$--$E_{\rm iso}$ and Combo correlations have been calibrated via OHD actual and machine-learned data, and again, based on a joint analysis with SNeIa and BAO, indications against a genuine cosmological constant have been found.
Analogously, in Ref.~\cite{Rezaeietal2020} different combinations of SNe Ia, quasar, and GRB data sets have been used for testing the $\Lambda$CDM model and dynamical dark energy parametrizations. It was found that GRB and quasar data sets were inconsistent with the flat $\Lambda$CDM model, in agreement with  Ref.~\cite{Lussoetal2019} for similar data.
In Ref.~\cite{Kumaretal2021} strong gravitational lensing data in conjunction with SNe Ia and GRBs have been considered and it has been found that the best-fit value of the spatial curvature parameter  favored a closed universe, although a flat universe can be accommodated at the $68\%$ confidence level.

On the other hand, some recent efforts have shown that the $E_{\rm p}$--$E_{\rm iso}$ and Combo correlations calibrated using better-established cosmological data --- such as SNe Ia or OHD measurements --- provide cosmological constraints that are consistent with the flat $\Lambda$CDM model, see Table~\ref{tab:recap}. In Ref.~\cite{2019MNRAS.486L..46A} an updated $E_{\rm p}$--$E_{\rm iso}$ correlation with $193$ GRBs and a calibration based on an interpolation of the OHD data set have been considered, leading to $\Omega_{m0}=0.397_{-0.039}^{+0.040}$ in a flat $\Lambda$CDM cosmology, though the value of the mass density is higher than the one established by Ref.~\cite{PlanckCollaboration2020}. In Ref.~\cite{Montieletal2021} the $E_{\rm p}$--$E_{\rm iso}$ correlation, calibrated with the latest OHD data set, has been jointly fit with CMB, BAO and SNe Ia data in a search for cosmological parameter constraints within the standard cosmological model, as well as in dynamical dark energy parametrizations, finding no evidence in favour of the alternatives to the $\Lambda$CDM model. Finally, by using the Combo correlation with $174$ GRBs calibrated in a semi-model independent way, in Ref.~\cite{2021ApJ...908..181M} it has been found: a) for a flat $\Lambda$CDM model $\Omega_{m0}=0.32^{+0.05}_{-0.05}$ and $\Omega_{m0}=0.22^{+0.04}_{-0.03}$ for the two values of the Hubble constant $H_0$ of Ref.~\cite{PlanckCollaboration2020} and Ref.~\cite{2019ApJ...876...85R}, respectively, and b) for a non-flat $\Lambda$CDM model $\Omega_{ m0}=0.34^{+0.08}_{-0.07}$ and $\Omega_\Lambda=0.91^{+0.22}_{-0.35}$ for the $H_0$ of Ref.~\cite{PlanckCollaboration2020}, and $\Omega_{m0}=0.24^{+0.06}_{-0.05}$ and $\Omega_\Lambda=1.01^{+0.15}_{-0.25}$ for the $H_0$ of Ref.~\cite{2019ApJ...876...85R}.
\begin{table*}
\centering
\setlength{\tabcolsep}{0.33em}
\renewcommand{\arraystretch}{1.5}
\begin{tabular}{l|c|ccccc}
\hline\hline
Correlation   &  Sample  &  $H_0$~(km~s$^{-1}$Mpc$^{-1}$)
              &  $\Omega_{\rm m0}$
              &  $\Omega_{\Lambda}$
              &  $\Omega_{\rm k0}$
              & Ref.\\
\hline
Amati ($193$) &  GRB+SNIa  &  $67.76\pm3.68^{a}$
              &  $0.397_{-0.039}^{+0.040}$
              &  $0.603_{-0.039}^{+0.040}$
              &  $0$
              & \cite{2019MNRAS.486L..46A}\\
\hline
Amati ($74$)  &  GRB+SNIa+BAO+CMB  &  $70.81\pm3.68{b}$
              &  $0.3180\pm0.0006$
              &  $0.6820\pm0.0006$
              &  $0$
              & \cite{Montieletal2021}\\
\hline
Combo ($174$) &  GRB  &  $74.03\pm1.42^{c}$
              &  $0.22^{+0.04}_{-0.03}$
              &  $0.78_{-0.04}^{+0.03}$
              &  $0$
              & \cite{2021ApJ...908..181M}\\
              &         &
              &  $0.24^{+0.06}_{-0.05}$
              &  $0.68^{+0.05}_{-0.05}$
              &  $-0.24^{+0.16}_{-0.25}$
              & \\
              &         &  $67.4\pm0.5^{d}$
              &  $0.32^{+0.05}_{-0.05}$
              &  $0.68^{+0.05}_{-0.05}$
              & $0$
              & \\
              &         &
              &  $0.34^{+0.08}_{-0.07}$
              &  $0.91^{+0.22}_{-0.35}$
              &  $-0.24^{+0.24}_{-0.35}$
              & \\
\hline
Amati ($118$) &  GRB+$H(z)$+BAO &  $68.544^{+0.871}_{-0.862}$
              &  $0.316\pm0.016$
              &  $0.684\pm0.016$
              &  $0$
              & \cite{2020MNRAS.499..391K}\\
              &  &  $67.499^{+2.281}_{-2.279}$
              &  $0.310\pm0.016$
              &  $0.639^{+0.072}_{-0.078}$
              &  $0.051^{+0.094}_{-0.088}$
              & \\
\hline
Amati ($118$) &  GRB+$H(z)$+BAO+QSO+HIIG &  $69.3\pm1.2$
              &  $0.313\pm0.013$
              &  $0.687\pm0.013$
              &  $0$
              & \cite{Caoetal2020}\\
\hline
\hline
\end{tabular}
\caption{Summary of some recent cosmological constraints obtained by using Amati and Combo correlations, with or without other well established cosmological probes, within the flat and non-flat $\Lambda$CDM models. The numbers near the correlations name indicate the size of the GRB sample. For details on the names of the other probes, see the text.\\
$^{a}$ \footnotesize{Inferred from the interpolation of the OHD data by using B\'ezier polynomials.}\\
$^{b}$ \footnotesize{Inferred from the interpolation of the OHD data with additional systematic errors \cite{Montieletal2021} by using B\'ezier polynomials.}\\
$^{c}$ \footnotesize{Value from Ref.~\cite{2019ApJ...876...85R}.}\\
$^{d}$ \footnotesize{Value from Ref.~\cite{Planck2018}.}}
\label{tab:recap}
\end{table*}

Again, by examining an uncalibrated $E_{\rm p}$--$E_{\rm iso}$ correlation built up from a sample of bright $Fermi$-LAT GRBs \citep{2019ApJ...887...13F} and another GRB sample with lower average fluence GRBs \citep{2016A&A...585A..68W}, in Ref.~\cite{2020MNRAS.499..391K} cosmological parameter constraints have been obtained in a number of cosmological models, concluding that current GRB data are not able to restrictively constrain cosmological parameters, and that cosmological parameter constraints from the more-reliable GRBs are consistent with those resulting from better-established cosmological probes.
In Ref.~\cite{Caoetal2021a}, a joint $H(z)$+BAO+quasar (QSO)+HII starburst galaxy (HIIG)+GRB fit determined $\Omega_{m0}=0.313\pm0.013$ in the flat $\Lambda$CDM model, consistency with a cosmological constant and zero spatial curvature, though mild dark energy dynamics or a little spatial curvature are not ruled out at all.

Mixing all together, the  cosmological results summarized above, obtained through GRB data, seem to be mutually inconsistent.
This reflects  all the efforts made so far to  employ GRB as distance indicators are still affected by a certain number of issues, as we outlined previously.

First of all, we recall that GRB correlations involve a number of observable quantities affected by the so-called circularity problem \citep{Kodama2008}, caused by having to compute the GRB correlations in a \emph{a priori} assumed background cosmological model, being  not fully model-independent \citep{Dainotti2008, SamushiaRatra2010, Bernardini2012, AmatiDellaValle2013, Wei2014, Izzo2015, Demianskietal2017a, Demianskietal2017b}. However, even uncalibrated GRB correlations, in principle free from the circularity issue, are not able to put stringent constraints on the cosmological parameters, though consistent with those resulting from better-established cosmological probes.
In addition, we recall all GRB correlations are characterized by large intrinsic dispersions, conceivably caused by unknown large systematic errors\footnote{Possibly including those associated with detector sensitivity, and the differences in estimated spectral parameters determined from measurements taken with different detectors or from different models.} \citep{Schaefer2007,2008MNRAS.391..411B,2009AIPC.1133..350N,2011MNRAS.418L.109G} in comparison to the case of better-established probes, such as BAO, OHD, and SNeIa, where many error sources have been better modeled. On the other hand, the influence of possible selection bias and evolution effects are currently debated \citep{Butler_2007,2008MNRAS.387..319G,2008MNRAS.391..639N,2009A&A...508..173A,2010PASJ...62.1495Y}. One may therefore conclude that the large intrinsic dispersions of GRB correlations could be a consequence of  yet undiscovered GRB intrinsic properties and/or a yet unidentified sub-class within the population of GRBs, analogously to SN populations.

\subsection{Applications of statistical analysis with GRBs}\label{10.2}

In this section, we describe a few applications of statistical analysis using GRBs. Clearly, we focus on a particular choice and, in principle, it is possible to work out different fits and/or experimental procedures. In particular, we here propose a calibration at the very beginning, adopting the most consolidate route to handle GRBs. Above we also described the non-calibration procedure that, for brevity, we do not report here.

A widely consolidate approach is based on sampling the original catalog by means of a Monte Carlo technique, essentially built up using Markov Chain Monte Carlo  simulations, that are sampled within the widest possible parameter space. Commonly, the most adopted algorithm is the \emph{Metropolis-Hastings}  and the standard approach to get limits uses the minimization of the total $\chi^2$ function. As we stated above, in this review, the idea of combining more than one sample is essential in order to refine cosmological bounds on the model parameters. Hereafter we denote with ${\bf x}$ the set of parameters and include in our analysis  SNe Ia and BAO data sets together with the calibrated GRB data. The former data have been obtained through calibrating the correlations. For the sake of brevity, as well as above, we only consider Amati, Ghirlanda, Yonetoku  and Combo correlations. In the specific case of our three samples, i.e., GRBs, SNeIa and BAO, we combine the chi square functions by
\begin{equation}
\chi^2_{\rm tot}=\chi^2_{\rm GRB}+\chi^2_{\rm SN}+\chi^2_{\rm BAO}\,,
\end{equation}
with the following recipe.
\begin{itemize}
\item[-]{\bf GRB $\chi^2$}. Here, we define
\begin{equation}
\label{chisquared}
\chi^2_{\rm GRB}=\sum_{i=1}^{N_{\rm GRB}}\left[\dfrac{\mu_{\rm GRB,i}^{\rm obs}-\mu_{\rm GRB}^{\rm th}\left({\bf x},z_i\right)}{\sigma_{\mu_{\rm GRB,i}}}\right]^2\,,
\end{equation}
where $N_{\rm GRB}$ and $\mu_{\rm GRB}^{\rm th}$ are the experimental and theoretical GRB distance moduli.
\item[-]{\bf SN $\chi^2$}. Here, by virtue of the above discussion concerning SN statistical analysis, we rewrite the chi square function in Eq.~\eqref{chisne1} by
\begin{equation}
\chi^2_{\rm SN}=\left(\Delta \mu_{\rm SN}- \mathcal{M}\mathbf{1} \right)^{\rm T} \mathbf{C}^{-1}
\left(\Delta\mu_{\rm SN}-\mathcal{M} \mathbf{1} \right)\,,
\end{equation}
where $\Delta\mu_{\rm SN}\equiv \mu_{\rm SN}-\mu_{\rm SN}^{\rm th}\left({\bf x},z_i\right)$ is the module of the vector of residuals, and $\mathbf{C}$  the covariance matrix.

In particular, we prompt the distance modulus for the most recent SN catalog, named \textit{Pantheon Sample}. This  represents the current largest SN sample consisting of $1048$ SNe Ia lying on $0.01<z<2.3$ \citep{2018ApJ...859..101S}. The corresponding magnitudes read
\begin{equation}
\mu_{\rm SN}=m_{\rm B}- \left(\mathcal{M}-\alpha \mathcal{X}_1+\beta \mathcal{C} -\Delta_{\rm M}-\Delta_{\rm B}\right)\ .
\end{equation}
Here, $\mathcal{M}$ and $m_{\rm B}$ are the $B$-band absolute and apparent magnitudes, respectively.
The above distance moduli also depend upon other quantities required to standardize/correct the light curves of SNe Ia. The quantities $\mathcal{X}_1$ and $\mathcal{C}$ are the light curve shape and color parameters, respectively, whereas $\alpha$ and $\beta$ are the coefficients of the luminosity-stretch and luminosity-color relationships, respectively. $\Delta_{\rm M}$ is a distance correction determined on host galaxy mass of SNe, while $\Delta_{\rm B}$ a distance correction that is built up from predicted biases determined by means of simulations.

By marginalizing over $\mathcal{M}$ through a flat prior, it is possible to demonstrate that SN uncertainties do not depend on $\mathcal{M}$ and this permits one to simplify the chi square function through
\begin{equation}
 \chi^2_{{\rm SN},\mathcal{M}} = a + \log \frac{e}{2 \pi} - \frac{b^2}{e}\,,
 \label{eqn:chimarg}
\end{equation}
where $a\equiv\Delta\vec{\mathbf{\mu} }_{\rm SN}^{T}\mathbf{C}^{-1}\Delta\vec{\mathbf{\mu} }_{\rm SN}$, $b\equiv\Delta\vec{\mathbf{\mu} }_{\rm SN}^{T}\mathbf{C}^{-1}\vec{\mathbf{1}} $, $e \equiv
\vec{\mathbf{1}}^T\mathbf{C}^{-1} \vec{\mathbf{1}}$.
\item[-]{\bf BAO $\chi^2$}. The chi square function for BAO data is given in Eqs.~\eqref{ABAO}--\eqref{chiBAO}.
\end{itemize}

Below we summarize a couple of statistical methods applied to GRB data to extract cosmological constraints.

\subsubsection{B\'ezier polynomials and cosmographic series}\label{10.2.1}

The first method utilizes GRB data, calibrated through the above B\'ezier polynomials, to extract cosmological constraints by means of cosmographic model-independent series and heal \emph{de facto} the circularity problem without postulating the model \emph{a priori} \cite{orlando3}.
This method has been implemented to Amati, Ghirlanda, Yonetoku and Combo GRB correlations in conjunction with SNe Ia and BAO data sets to get more stable and narrow constraints.
We considered the most recent approaches to cosmography, comparing among them Taylor expansions with $z$ and $y_2$ series, and Pad\'e polynomials.
Two hierarchies have been considered: \textit{hierarchy 1}, up to $j_0$, and \textit{hierarchy 2}, up to $s_0$.

Reasonable results have been found up for both hierarchies through several MCMC fits showing possible matching with the standard paradigm (see Tables \ref{tab:summarytaylor}--\ref{tab:summarychisquare}). Moreover, we only partially alleviated the tension on local $H_0$ measurements as hierarchy 2 is considered.
Taylor outcomes are quite stable within each hierarchy, as portrayed by the results in Table~\ref{tab:summarytaylor}, and work well with Amati, Ghirlanda and Yonetoku correlations in the sense that the corresponding numerical outcomes are consistent within $1$--$\sigma$ with previous findings. Again, this suggests a spatially flat $\Lambda$CDM paradigm as statistically favored model, with mass density parameter $\Omega_m=2(1+q_0)/3\sim0.3$ for Combo correlation, whereas the other correlations seem to indicate smaller values.

The auxiliary $y_2$ variable is not enough stable than Taylor expansions. It  enlarges significantly $h_0$, see e.g. Tab.~\ref{tab:summary2} and the overall results are however quite unpredictive at the level of hierarchy 1.
Moreover, Pad\'e fits seem to improve the quality of Taylor expansion hierarchy 1, as expected by construction. This is  particularly evident for Combo and Yonetoku correlations, while for Amati and Ghirlanda correlations is not. It is worth noticing that to go further jerk term implies $\geq (3,1)$, leading to higher orders than  $P_{3,1}$, quite unconstrained at higher redshift domains.

Quite surprisingly, our findings summarized in Tables \ref{tab:summarytaylor}--\ref{tab:summarychisquare} show that the $\Lambda$CDM model is not fully confirmed using GRBs. Although this can be an indication that more refined analyses are needful, as GRBs are involved, simple indications seem to be against a genuine cosmological constant \cite{PlanckCollaboration2020} and may be interpreted either with a barotropic dark energy contribution or with the need of non-zero spatial curvature \cite{orlando3}.
Nevertheless, at this stage, our findings are in line with recent claims on tensions with the $\Lambda$CDM model \citep{2019arXiv191101681Y,Lussoetal2019,2019NatAs...3..272R}.
\begin{table*}
\centering
\setlength{\tabcolsep}{0.22em}
\renewcommand{\arraystretch}{1.5}
\begin{tabular}{l|ccc|cccc}
\hline\hline
       & \multicolumn{7}{c}{\bf Taylor fits}        \\
\cline{2-8}
 & \multicolumn{3}{c}{\emph{Hierarchy} $1$}
            &  \multicolumn{4}{c}{\emph{Hierarchy} $2$}\\
\cline{2-8}
            &  $h_0$
            &  $q_0$
            &  $j_0$
            &  $h_0$
            &  $q_0$
            &  $j_0$
            &  $s_0$\\
\hline
A           &  $0.740_{-0.006\,(-0.013)}^{+0.005\,(+0.010)}$
            &  $-0.68_{-0.02\,(-0.04)}^{+0.03\,(+0.06)}$
            &  $0.77_{-0.10\,(-0.20)}^{+0.08\,(+0.16)}$
            &  $0.700_{-0.008\,(-0.015)}^{+0.007\,(+0.014)}$
            &  $-0.51_{-0.01\,(-0.02)}^{+0.02\,(+0.03)}$
            &  $0.71_{-0.05\,(-0.10)}^{+0.06\,(+0.12)}$
            &  $-0.36_{-0.10\,(-0.20)}^{+0.05\,(+0.13)}$\\
G &  $0.716_{-0.006\,(-0.014)}^{+0.006\,(+0.013)}$
            &  $-0.63_{-0.03\,(-0.05)}^{+0.03\,(+0.06)}$
            &  $0.76_{-0.09\,(-0.18)}^{+0.09\,(+0.17)}$
            &  $0.691_{-0.007\,(-0.015)}^{+0.008\,(+0.016)}$
            &  $-0.50_{-0.02\,(-0.05)}^{+0.02\,(+0.05)}$
            &  $0.64_{-0.10\,(-0.19)}^{+0.06\,(+0.15)}$
            &  $-0.42_{-0.08\,(-0.16)}^{+0.10\,(+0.17)}$\\
Y &  $0.737_{-0.008\,(-0.015)}^{+0.008\,(+0.014)}$
            &  $-0.73_{-0.01\,(-0.04)}^{+0.03\,(+0.06)}$
            &  $0.88_{-0.13\,(-0.23)}^{+0.02\,(+0.13)}$
            &  $0.695_{-0.008\,(-0.015)}^{+0.007\,(+0.014)}$
            &  $-0.54_{-0.01\,(-0.03)}^{+0.02\,(+0.04)}$
            &  $0.70_{-0.05\,(-0.11)}^{+0.07\,(+0.13)}$
            &  $-0.36_{-0.09\,(-0.18)}^{+0.08\,(+0.16)}$\\
C &  $0.706_{-0.007\,(-0.013)}^{+0.007\,(+0.013)}$
            &  $-0.59_{-0.03\,(-0.06)}^{+0.03\,(+0.07)}$
            &  $0.72_{-0.10\,(-0.18)}^{+0.09\,(+0.18)}$
            &  $0.693_{-0.009\,(-0.015)}^{+0.006\,(+0.014)}$
            &  $-0.52_{-0.01\,(-0.03)}^{+0.02\,(+0.05)}$
            &  $0.73_{-0.09\,(-0.15)}^{+0.06\,(+0.13)}$
            &  $-0.38_{-0.10\,(-0.19)}^{+0.06\,(+0.14)}$\\
\hline
\hline
\end{tabular}
\caption{Cosmographic best fits and $1$--$\sigma$ ($2$--$\sigma$) errors from Taylor expansions labeled as \textit{hierarchy 1} ($h_0$, $q_0$, $j_0$) and \textit{hierarchy 2} ($h_0$, $q_0$, $j_0$, $s_0$). Letters A, G, Y and C indicate Amati, Ghirlanda, Yonetoku and Combo correlations, respectively.}
\label{tab:summarytaylor}
\end{table*}
\begin{table*}
\centering
\setlength{\tabcolsep}{0.22em}
\renewcommand{\arraystretch}{1.5}
\begin{tabular}{l|ccc|cccc}
\hline\hline
 & \multicolumn{7}{c}{{\bf $y_2$ fits}}\\
\cline{2-8}
 &  \multicolumn{3}{c}{\emph{Hierarchy} $1$}
            &  \multicolumn{4}{c}{\emph{Hierarchy} $2$}\\
\cline{2-8}
            &  $h_0$
            &  $q_0$
            &  $j_0$
            &  $h_0$
            &  $q_0$
            &  $j_0$
            &  $s_0$\\
\hline
A &  $0.76_{-0.01\,(-0.02)}^{+0.01\,(+0.02)}$
            &  $-1.35_{-0.04\,(-0.08)}^{+0.05\,(+0.09)}$
            &  $3.85_{-0.28\,(-0.50)}^{+0.25\,(+0.49}$
            &  $0.78_{-0.01\,(-0.02)}^{+0.01\,(+0.02)}$
            &  $-0.53_{-0.04\,(-0.10)}^{+0.06\,(+0.11)}$
            &  $-2.52_{-0.42\,(-0.78)}^{+0.31\,(+0.71)}$
            &  $-4.41_{-0.58\,(-1.29)}^{+1.00\,(+1.82)}$\\
G &  $0.75_{-0.01\,(-0.02)}^{+0.01\,(+0.02)}$
            &  $-1.04_{-0.05\,(-0.10)}^{+0.05\,(+0.10)}$
            &  $2.40_{-0.23\,(-0.47)}^{+0.24\,(+0.47)}$
            &  $0.74_{-0.01\,(-0.02)}^{+0.01\,(+0.02)}$
            &  $-0.45_{-0.06\,(-0.13)}^{+0.10\,(+0.17)}$
            &  $-2.17_{-0.61\,(-1.14)}^{+0.42\,(+0.96)}$
            &  $-3.08_{-0.56\,(-1.28)}^{+1.41\,(+2.68)}$\\
Y &  $0.75_{-0.01\,(-0.02)}^{+0.01\,(+0.02)}$
            &  $-1.05_{-0.04\,(-0.09)}^{+0.05\,(+0.10)}$
            &  $2.47_{-0.23\,(-0.49)}^{+0.22\,(0.50)}$
            &  $0.74_{-0.01\,(-0.02)}^{+0.01\,(+0.02)}$
            &  $-0.43_{-0.10\,(-0.21)}^{+0.03\,(+0.08)}$
            &  $-2.19_{-0.32\,(-0.63)}^{+0.62\,(+1.62)}$
            &  $-2.70_{-1.04\,(-1.51)}^{+0.37\,(+0.86)}$\\
C &  $0.75_{-0.01\,(-0.02)}^{+0.01\,(+0.02)}$
            &  $-1.01_{-0.05\,(-0.09)}^{+0.04\,(+0.9)}$
            &  $2.29_{-0.20\,(-0.40)}^{+0.23\,(+0.44)}$
            &  $0.74_{-0.01\,(-0.02)}^{+0.01\,(+0.02)}$
            &  $-0.43_{-0.09\,(-0.17)}^{+0.06\,(+0.16)}$
            &  $-2.19_{-0.38\,(-1.03)}^{+0.64\,(+1.19)}$
            &  $-2.79_{-0.82\,(-1.50)}^{+0.90\,(+2.59)}$\\
\hline
\hline
\end{tabular}
\caption{Cosmographic best fits and $1$--$\sigma$ ($2$--$\sigma$) errors from expansions with $y_2$ labeled as \textit{hierarchy 1} ($h_0$, $q_0$, $j_0$) and \textit{hierarchy 2} ($h_0$, $q_0$, $j_0$, $s_0$). Letters A, G, Y and C indicate Amati, Ghirlanda, Yonetoku and Combo correlations, respectively.}
\label{tab:summary2}
\end{table*}
\begin{table}
\centering
\setlength{\tabcolsep}{0.22em}
\renewcommand{\arraystretch}{1.5}
\begin{tabular}{l|ccc}
\hline\hline
& \multicolumn{3}{c}{{\bf Pad\'e fits}}\\
\cline{2-4}
            &  \multicolumn{3}{c}{\emph{Hierarchy} $1$}\\
\cline{2-4}
            &  $h_0$
            &  $q_0$
            &  $j_0$\\
\hline
A &  $0.70_{-0.02\,(-0.03)}^{+0.01\,(+0.03)}$
            &  $-0.33_{-0.03\,(-0.08)}^{+0.05\,(+0.09)}$
            &  $0.240_{-0.010\,(-0.020)}^{+0.010\,(+0.020)}$\\
G &  $0.70_{-0.01\,(-0.02)}^{+0.02\,(+0.03)}$
            &  $-0.31_{-0.05\,(-0.09)}^{+0.02\,(+0.06)}$
            &  $0.235_{-0.002\,(-0.006)}^{+0.013\,(+0.027)}$\\
Y &
$0.68_{-0.01\,(-0.02)}^{+0.01\,(+0.02)}$
            &  $-0.32_{-0.04\,(-0.07)}^{+0.02\,(+0.06)}$
            &  $0.240_{-0.005\,(-0.010)}^{+0.010\,(+0.021)}$\\
C &  $0.68_{-0.01\,(-0.02)}^{+0.01\,(+0.02)}$
            &  $-0.33_{-0.03\,(-0.06)}^{+0.03\,(+0.06)}$
            &  $0.244_{-0.006\,(-0.012)}^{+0.009\,(+0.019)}$\\
\hline
\hline
\end{tabular}
\caption{Cosmographic best fits and $1$--$\sigma$ ($2$--$\sigma$) errors from Pad\'e expansions labeled as \textit{hierarchy 1}.  Letters A, G, Y and C indicate Amati, Ghirlanda, Yonetoku and Combo correlations, respectively.}
\label{tab:summarypade}
\end{table}

\begin{table}
\centering
\setlength{\tabcolsep}{0.33em}
\renewcommand{\arraystretch}{1.5}
\begin{tabular}{l|c|c|ccc}
\hline\hline
Sample        &
DoF           &
Hierarchy & \multicolumn{3}{c}{Approximant $\chi^2$}\\
\cline{4-6}
            &
            &
            &  Taylor
            &  Function $y_2$
            &  Pad\'e $P_{2,1}$\\
\hline
Combo
            &  $1113$
            &  $1$
            &  $1116.84$
            &  $1230.71$
            &  $1113.77$\\
            &  $1112$
            &  $2$
            &  $1089.25$
            &  $1160.04$
            &\\
\hline
Ghirlanda
            &  $1080$
            &  $1$
            &  $1120.19$
            &  $1271.92$
            &  $2203.16$\\
            &  $1079$
            &  $2$
            &  $1075.01$
            &  $1184.42$
            &\\
\hline
Yonetoku
            &  $1154$
            &  $1$
            &  $1235.08$
            &  $1350.27$
            &  $1178.07$\\
            &  $1153$
            &  $2$
            &  $1147.72$
            &  $1227.25$
            &\\
            \hline
Amati
            &  $1246$
            &  $1$
            &  $2334.35$
            &  $2818.25$
            &  $2202.75$\\
            &  $1245$
            &  $2$
            &  $2174.13$
            &  $2539.98$
            &\\
\hline
\hline
\end{tabular}
\caption{$\chi^2$ values of the cosmographic fits performed over the considered approximants. For each GRB correlations the number of degrees of freedom (DoF) and the considered hierarchy are reported. Correlations are sorted for increasing values of the ratio $\chi^2/$DoF with respect to the Taylor hierarchy $1$ expansion.}
\label{tab:summarychisquare}
\end{table}

\subsubsection{B\'ezier polynomials and $\Lambda$CDM and $\omega$CDM cosmological models}\label{10.2.2}

In a second method here summarized, the circularity problem affecting GRBs is again overcome by using B\'ezier polynomials to calibrate, in this case, the Amati correlation alone \cite{2019MNRAS.486L..46A}.
Unlike the previous method, now GRB data are utilized in conjuction with the SNe Ia JLA data set alone \cite{2014A&A...568A..22B} and employed to explicitly constrain two different cosmological scenarios: the concordance $\Lambda$CDM model and the $\omega$CDM model, with the dark energy equation of state parameter free to vary \cite{2019MNRAS.486L..46A}.

In the Monte Carlo integration, through the Metropolis-Hastings algorithm, $H_0$ has been fixed to the best-fit value obtained from the model-independent analysis over OHD data, i.e. $H_0=67.76$~km~s$^{-1}$~Mpc$^{-1}$.
The results for $\Omega_m$ and $w$ (see Table \ref{tab:results}) agree with previous findings making use of GRBs.
The statistical performance of the models under study has been evaluated through the  Akaike information criterion (AIC) criterion \cite{1974ITAC...19..716A}
\begin{equation*}
\text{AIC}\equiv\ 2p -2\ln \mathcal{L}_{max}\,,
\end{equation*}
where $p$ is the number of free parameters in the model and $\mathcal{L}_{max}$ is the maximum probability function calculated at the best-fit point, and the  deviance information criterion (DIC) criterion \citep{Kunz2006}
\begin{equation*}
\text{DIC}\equiv2p_{eff}-2\ln \mathcal{L}_{max}\ ,
\end{equation*}
where $p_{eff}=\langle-2\ln\mathcal{L}\rangle+2\ln\mathcal{L}_{max}$ is the number of parameters that a data set can effectively constrain.\footnote{Here, the brackets indicate the average over the posterior distribution.}
The best model is the one that minimizes the AIC and DIC values. Unlike the AIC criterion, the DIC statistics does not penalize for the total number of free parameters of the model, but only for those which are constrained by the data \citep{Liddle2007}.
Differently from the previous approach, we found that the $\Lambda$CDM model is preferred with respect to the minimal $\omega$CDM extension (see Table~\ref{tab:results}) and then conclude that no modifications of the standard paradigm are expected as intermediate redshifts are involved.
However, future efforts dedicated to the use of our new technique to fix refined constraints over dynamical dark energy models are encouraged in order to fix the apparent dichotomy in the results of the two described methods.
\begin{table}
\centering
\small
\setlength{\tabcolsep}{1em}
\renewcommand{\arraystretch}{1.2}
\caption{95\% confidence level results of the MCMC analysis for the SN+GRB data. The AIC and DIC differences are intended with respect to the $\Lambda$CDM model.}
\begin{tabular}{c c c c c}
\hline
\hline
Model & $\omega$ & $\Omega_{m}$ & $\Delta$AIC & $\Delta$DIC\\
\hline
$\Lambda$CDM &  -1& $0.397^{+0.040}_{-0.039}$ & 0 & 0 \\
$w$CDM & $-0.86^{+0.36}_{-0.38}$ & $0.34^{+0.13}_{-0.15}$ & $1.44$ & $1.24$ \\
\hline
\hline
\end{tabular}
\label{tab:results}
\end{table}

\subsection{The role of spatial curvature}\label{sec:5}

An update  sample of GRBs has been developed in 2020, in which the Combo relation extracts bounds on the spatial curvature with no other probes \cite{2021ApJ...908..181M}, differently from previous attempts that commonly assume a spatially flat background, with the inclusion of SNe Ia and BAO.

The way in which the Combo relation is calibrated is without OHD data set, but rather invoking two step methods. In this picture, we assume \cite{Izzo2015}

\begin{itemize}
\item[{\bf I.}]{the terms $k_1$ and $\sigma_{\rm k}$ are got from small GRB sub-samples with almost the same redshift;}
\item[{\bf II.}]{$k_0$ is determined from the use of SNe Ia limited to the lowest redshift of the GRBs of the Combo data set, in which the calibration of SNe Ia is negligible \cite{Izzo2015}.}
\end{itemize}

\begin{table}
\centering
\caption{Best-fit parameters of the seven sub-samples at average redshift $\langle z\rangle$: the slope $q_{\rm 1,z}$, the normalization $\log F_{\rm 0,z}$, and the extrascatter $\sigma_{\rm q,z}$ are shown.}
\begin{tabular}{c|ccc}
\hline\hline
$\langle z\rangle$  &   $k_{\rm 1,z}$       &  $\log \left[F_{\rm 0,z}/\left({\rm erg~cm^{-2}s^{-1}}\right)\right]$    & $\sigma_{\rm k,z}$ \\
\hline
\hline
$0.54\pm0.01$	&	$0.81\pm0.49$	&	$-7.38\pm1.08$	                                    &	$0.29\pm0.10$	\\
$1.18\pm0.07$	&	$0.83\pm0.31$	&	$-7.93\pm0.77$	                                    &	$0.26\pm0.08$	\\
$1.46\pm0.05$	&	$0.80\pm0.32$	&	$-8.29\pm0.79$	                                    &	$0.26\pm0.08$	\\
$1.70\pm0.06$	&	$0.94\pm0.32$	&	$-8.92\pm0.84$	                                    &	$0.19\pm0.08$	\\
$2.05\pm0.05$	&	$1.05\pm0.32$	&	$-9.44\pm0.90$							            & $0.34\pm0.08$ \\
$2.27\pm0.07$	&	$0.78\pm0.20$	&	$-8.57\pm0.54$							           & $0.16\pm0.06$	\\
$2.69\pm0.08$	&	$1.02\pm0.34$	&	$-9.38\pm0.87$							           & $0.34\pm0.10$ \\
\hline												\hline
\end{tabular}
\label{tab:no3}
\end{table}

GRB sub-samples with the same $z$ are chosen among those ones providing well constrained best-fit parameters.

Considering that  in each sub-sample the GRB luminosity distances $d_{\rm L}$ are quite the same, we employ the rest-frame $0.3$--$10$ keV energy flux $F_0$. This improves the dependence on the model and enables one to  render our procedure cosmology-independent. In the specific example, here reported, seven sub-samples have been suggested and the best fits reported in Tab.~\ref{tab:no3}, showing no evident trends with $z$ within the errors.
This technique is used for Combo relation since previous results suggest its advantages in the fitting strategies above described\footnote{By construction, these  sub-samples exhibit the same correlation with the same $q_1$ and $\sigma_{\rm q}$, but involving different normalizations.}.
One can perform a simultaneous fit of the above sub-samples with the same $k_1$ and $\sigma_{\rm k}$ implying $k_1=0.90\pm0.13$ and $\sigma_{\rm k}=0.28\pm0.03$.

The calibration of $k_0$ is performed by means of the nearest couple of GRBs of the employed sample with the same redshift. In particular, it is possible to take  GRB~130702A at $z=0.145$ and GRB~161219B at $z=0.1475$.
Then, $\mu_{\rm C}^{\rm obs}$ can be replaced via its average distance modulus $\langle\mu_{\rm SNIa}\rangle=39.21\pm0.24$, determined by SNe Ia with the same $z$ of the above two GRBs, considering the bound over $k_1$ and the values of $F_0$, $E_{\rm p}$, $\tau$, and $\alpha$ for the two GRBs adopted throughout the computation. The computed value is $k_0=49.54\pm0.21$.

At this stage, comparing between GRB distance moduli $\mu_{\rm C}^{\rm obs}$, with uncertainties $\sigma\mu_{\rm C}^{\rm obs}$,  with theoretical expectations, it is possible to get constraints over background cosmologies. In particular, for a non-flat $\Lambda$CDM model, i.e., the simplest scenario to work with, we write
\begin{equation}
\label{eq:no5}
d_{\rm L} = \frac{c}{H_0}\frac{(1+z)}{\sqrt{| \Omega_k |}} {\rm sinn}\left(\int_0^z \frac{\sqrt{| \Omega_k |} dz^\prime}{\sqrt{\Omega_m(1+z)^3+\Omega_\Lambda+\Omega_k(1+z)^2}}\right)\ ,
\end{equation}
and we compute numerical bounds once $H_0$ is marginalized\footnote{Constraints come from  $H_0=(67.4\pm0.5)$~km~s$^{-1}$Mpc$^{-1}$ \cite{Planck2018} and $H_0=(74.03\pm1.42)$~km~s$^{-1}$Mpc$^{-1}$ \cite{2019ApJ...876...85R}.} as in Tab.~\ref{tab:no5}.

In particular, slightly larger estimations on matter density are got from
GRBs, i.e.,  $\Omega_m=0.32^{+0.05}_{-0.05}$ with $H_0$ of Ref.~\cite{Planck2018} and the opposite, i.e.,  $\Omega_m=0.22^{+0.04}_{-0.03}$, for the $H_0$ of Ref.~\cite{2019ApJ...876...85R}. Analogous results, i.e., compatible with the flat case, are computed using the non-flat $\Lambda$CDM model.

\begin{table}
\setlength{\tabcolsep}{0.75em}
\renewcommand{\arraystretch}{1.2}
\centering
\caption{Best-fit parameters with $1$--$\sigma$ uncertainties for the various cosmological cases discussed in this work. The last column lists the values of the $\chi^2$. $H_0$ is fixed to the values given by Ref.~\cite{Planck2018} and Ref.~\cite{2019ApJ...876...85R}, respectively the Planck and Riess expectation values.}
\small
\begin{tabular}{lcccc}
\hline\hline
$H_0$ &  $\Omega_m$       &  $\Omega_{\Lambda}$  & $\Omega_k$  &  $\chi^2$         \\
\hline
\hline
\multicolumn{5}{c}{Flat $\Lambda$CDM (${\rm DOF}=173$)}\\
\cline{1-5}
P18	   &   $0.32^{+0.05}_{-0.05}$   &  $0.68^{+0.05}_{-0.05}$   & $0$    &  $165.54$  \\
R19   &   $0.22^{+0.04}_{-0.03}$   &  $0.78^{+0.04}_{-0.03}$   & $0$    &  $171.32$  \\
\hline
\hline
\multicolumn{5}{c}{$\Lambda$CDM (${\rm DOF}=172$)} \\
\cline{1-5}
P18	   &   $0.34^{+0.08}_{-0.07}$   &  $0.91^{+0.22}_{-0.35}$   & $-0.24^{+0.24}_{-0.35}$     &  $164.38$  \\
R19   &   $0.24^{+0.06}_{-0.05}$   &  $1.01^{+0.15}_{-0.25}$   & $-0.24^{+0.16}_{-0.25}$   &  $169.23$ \\
\hline
\hline
\end{tabular}
\label{tab:no5}
\end{table}

\section{Further application of GRBs as probes of the high-redshift Universe}

Above we faced some the standard statistical methods associated with GRBs and we propose a few applications with the corresponding experimental bounds. Now we shortly review other phenomenological applications of GRBs that can shed light on the high-redshift Universe.

\subsection{Star formation rate  from GRBs}\label{sec:SFR}

LGRBs are likely associated with core-collapse SNe
\cite{Stanek2003,Hjorth2003}. By virtue of this fact,
GRB-SN associations may provide a complementary technique that measures high-redshift star formation rate
\cite{Totani97,Wijers98,LambReichart,Porciani01,Bromm02}. Again a problem related to calibration occurs, i.e. how to calibrate GRB event rate with star formation rate. Moreover, the luminosity function is not known {\it a priori} and its reconstruction depends upon the particular model selected for the analysis
\cite{Wanderman2010,Cao11,Tan13,2016A&A...587A..40P}. The typical functional structures for the luminosity functions are i) a broken power law \cite{Virgili2011}, ii) a single power law with an exponential cut-off at low luminosities \cite{Cao11}.
Reconstructed SFRs from GRBs are typically larger than those from other observations \cite{2015NewAR..67....1W}. The reason behind this apparent inconsistency may reside in the fact that usually SFR at high-$z$ is got from the observations of the brightest galaxies, whereas GRBs, in view of their high luminosities, may help in detecting faint galaxies at high-$z$ otherwise unobserved \cite{Wang13,2015NewAR..67....1W}.

\subsection{High-redshift GRB rate excess}

Although appealing, the above developments do not show why GRBs do not  follow
the star formation history being enhanced by hidden high-redshift mechanisms
\citep{Le07,Salvaterra07,Kistler08,Wang09,Robertson12,Wang13}.
In particular, the star formation rate at high-redshift, namely $z>6$, appears too large if confronted to star formation rate got from  high-redshift
galaxy surveys \cite{Bouwens2009}.

A natural origin of the high-redshift GRB rate excess can be found in the
metallicity evolution, as LGRBs seem to prefer low-metallicity environment, as supported by recent studies that favor such a requirement\footnote{There is uncertainty on measuring GRB metallicity at high-redshifts, as due to chemical inhomogeneity for example
\citep{Levesque10,Niino11}. Thus, this approach cannot be seen as definitive.
}.
Typical mass bounds on stars
suggest $>30M_\odot$, being responsible for BH remnants.

\subsection{Gravitationally-lensed GRBs}

Gravitationally lensed GRBs (GLGRBs) have been proposed in Ref.~\cite{Paczynski1986}, where it was speculated that such a phenomenon would result in multiple light curves detected at different times, as due to the different light paths of the produced multiple GRB images.

Quests for GLGRBs were mostly based on strong lensing\footnote{In the case of strong lensing, the time delay between the images is larger than the duration of the burst.} and similarities among GRB light curves with identical spectra and close locations in the sky, as primary search citeria \cite{1996AIPC..384..487M,2011AIPC.1358...17D,2014SCPMA..57.1592L,2019ApJ...871..121H,2020ApJ...897..178A}.
However, such searches led to null results, possibly due to Poisson noise uncertainties, affecting GRB light curves specially at low signal-to-noise ratios, which may have introduced large differences between the lensed GRB images \cite{2019ApJ...871..121H}. On the other hand, some GLGRB could exhibit time delays shorter than (or comparable to) the burst duration, hence leading to unresolved (or locally separated) peaks separated by the time delay \cite{2003CosRe..41..141O,2006ApJ...650..252H,2018PhRvD..98l3523J}.

Several searches have been performed in the literature, resulting in a few or null candidates, based on different techniques and lens models, such as globular cluster with mass of $\approx 10^5$--$10^7$ ~$M_{\odot}$ \cite{2003CosRe..41..141O}, Population III stars with a mass range of $10^2$--$10^3$~$M_{\odot}$ \cite{2006ApJ...650..252H}, diverse objects with a mass range of $10^2$--$10^7$~$ M_{\odot}$ \cite{2021NatAs...5..560P}, or a supermassive BH with a mass in the range of $\approx 10^5$--$10^7$~$M_{\odot}$ \cite{2021arXiv210500585K}.

Considering models where the lens is a supermassive BH \cite{2021NatAs...5..560P,2021arXiv210500585K}, GLGRB candidates can provide an opportunity to estimate the number density of massive compact objects at cosmological distances by calculating the rate of GRB lensing.
Assuming such supermassive BH lenses with mass $\approx10^6$~$M_\odot$ as dark matter compact objects \cite{1999ApJ...512L..13M}, the density parameter of BHs is $\Omega_{\rm BH}=0.007 \pm 0.004$, or equivalently $\Omega_{BH}/\Omega_M=0.027 \pm 0.016$ \cite{2021arXiv210500585K}.
In this respect, finding more GLGRBs candidates from supermassive BH may enhance our understanding of the matter content of the Universe.

\section{Final outlooks}\label{sec:6}

In this review, we outlined some of the most recent developments toward GRB physics, properties and their applications to cosmology. In particular, the review is structured into two main parts.

In the first part, we discussed the basic demands of GRBs, including their main observable quantities, their classification scheme into LGRBs and SGRBs and the corresponding issues, emphasizing new possible ways of classification, still object of speculation. Afterwards, we gave emphasis on GRB progenitors and on their fundamental microphysics, in view of the experimental evidences characterizing prompt and afterglow emissions, etc. LGRB connections with SNe have been explored as well along with SGRB matching with KNe and GWs. Great emphasis has been devoted to portray the standard GRB formation, working with the well consolidated \emph{fireball model}. Particle acceleration and radiative processes, predicted in such a picture, have been largely reported, with particular concern on observable signatures and on the standard model frontiers. Even in this part, we illustrated that GRBs cannot be contemplated as genuine standard candles, since there is no consensus toward their internal processes that we depicted throughout the manuscript. Accordingly, their luminosity cannot be easily put in relation with their redshifts as, for instance, one does for SNe.

For these aspects and for the overall limitations above described, we developed in the second part considerable cosmological applications of GRB physics. We tried to standardize GRBs by means of the most recent techniques and accentuated GRBs are essential to reconcile small with intermediate redshift domains, opening new scenarios toward our universe comprehension. In this respect, we featured how GRBs could be used as  complementary and outstanding probes to trace dark energy's evolution in support of other indicators, e.g. SNeIa, BAO, CMB, Hubble differential data, etc. Thereby we have shown a few statistical treatments related to Bayesian analysis in cosmology, able to combine GRBs with other catalogs of data,  reporting the most recent cosmological constraints on dark energy models. To do so, we expounded the bristly circularity problem, burdening GRBs in cosmological set ups. In particular, we also changed perspective, showing how to avoid calibration, {\it i.e.}, how not to employ the circularity. We confronted the two methods and checked which departures could be expected from the standard cosmological model through the use of GRBs in both the cases. Details on error propagation and GRB systematics have been discussed for several cosmic GRB correlations. Model dependent and independent techniques of calibrations have been likewise portrayed.

Perspectives about GRB developments will be based on clarifying the overall issues raised in this review. In particular, it is of utmost importance to shed light on how to standardize GRBs, in view of a likely self-consistent evolutionary paradigm, so far missing. With this recipe, we expect in the incoming years to improve GRB use in cosmology and get rid of circularity and greatly reduce the systematics and all the other issues that affect GRB data and challenge their use in cosmology. In particular, some models akin to those characterizing other cosmic indicators will spell out how to describe \emph{in toto} GRB physics and evolution.


\begin{acknowledgments}

The authors warmly thank Lorenzo Amati, Salvatore Capozziello, Massimo della Valle, Peter K.S. Dunsby, Luca Izzo,  Hernando Quevedo and Bharat Ratra for several discussions on GRBs and their applications to cosmology. The authors acknowledge the Ministry of Education and Science of the Republic of Kazakhstan, Grant: IRN AP08052311.
\end{acknowledgments}

\vspace{6pt}





%

\end{document}